\documentclass[12pt]{article}
\usepackage{a4wide}
\usepackage{amssymb}
\usepackage{amsmath}
\usepackage{hyperref}
\usepackage{graphicx}
\begin{document}
{\renewcommand{\thefootnote}{\fnsymbol{footnote}}
\begin{center}
{\LARGE  Canonical description of cosmological backreaction }\\
\vspace{1.5em}
Martin Bojowald\footnote{e-mail address: {\tt bojowald@gravity.psu.edu}}
and Ding Ding\footnote{e-mail address: {\tt dud79@psu.edu}}
\\
\vspace{0.5em}
Institute for Gravitation and the Cosmos,\\
The Pennsylvania State
University,\\
104 Davey Lab, University Park, PA 16802, USA\\
\vspace{1.5em}
\end{center}
}

\setcounter{footnote}{0}

\begin{abstract}
  Canonical methods of quasiclassical dynamics make it possible to go beyond a
  strict background approximation for cosmological perturbations by including
  independent fields such as correlation degrees of freedom. New models are
  introduced and analyzed here for cosmological dynamics in the presence of
  quantum correlations between background and perturbations, as well as
  cross-correlations between different modes of a quantum field. Evolution
  equations for moments of a perturbation state reveal conditions required for
  inhomogeneity to build up out of an initial vacuum. A crucial role is played
  by quantum non-locality, formulated by canonical methods as an equivalent
  local theory with non-classical degrees of freedom given by moments of a
  quantum state.
\end{abstract}

\section{Introduction}

The backreaction problem in cosmology is concerned with the question of how an
evolving or fluctuating quantum field in an expanding universe might affect
the expansion rate of the background geometry on which it is defined. In the
absence of a complete and consistent quantum theory of gravity or cosmology,
the background is mainly treated as a classical system. The backreaction
problem is therefore relevant not only for observational and conceptual
questions in cosmology (see for instance
\cite{BackreactionStochastic,BackreactionPreheating} for recent contributions),
it also offers general lessons for the classical-quantum correspondence
\cite{Halliwell2,Halliwell3,CosmoDecoh,QuantClassNonVac,QuantClassCosmo,QuantClassAnalog,FermionDecoherence,BKKM99,InflStruc,PointerInflation,CollapseStruc1,CollapseStruc2,CollapseStruc3}
and the validity of hybrid schemes in which only one part of an interacting
system is quantum
\cite{ClassQuantMeas,HybridChem,HybridQuant,HybridMadelung}. In this paper, we
apply recent advances in canonical descriptions of semiclassical expansions to
a system that has been used recently as a cosmological model \cite{CQC}, to
what turns out to be the leading semiclassical order of our systematic scheme.

We will use this systematic extension to construct new models for
early-universe cosmology which may be interpreted as being of hybrid
type. While we will not directly address the classical-to-quantum transition,
we will be able to shed new light on the related homogeneity problem of
inflationary structure formation \cite{InflStruc}, which states that it should
be impossible for translation-invariant quantum evolution to generate
inhomogeneity out of an initial homogeneous vacuum state. We will enter a
crucial new ingredient into this discussion, given by quantum non-locality
which is able to imply a gradual process of transfer of super-horizon
inhomogeneity --- suggested to occur generically according to the
Belinskii--Khalatnikov--Lifshitz (BKL) scenario \cite{BKL}) --- down to
smaller distance scales and eventually into the cosmological horizon. As an
implication of quantum non-locality, this process may happen even in the
quantization of a classical covariant theory. As we will observe for the first
time in an application to quantum field theory, canonical moment methods give
efficient access to spatial non-locality. (As has been known for some time,
moment methods allow one to describe quantum non-locality in time by foregoing
an adiabatic approximation of usual effective potentials.)

The classical Hamiltonian of the model used in \cite{CQC} is given by
\begin{equation} \label{H}
	H=\frac{1}{2}p_x^2+\frac{1}{2}p_z^2-ax+\frac{1}{2}(\omega^2+\lambda
        x^2)z^2 \, 
\end{equation}
with two degrees of freedom $x$ and $z$, where $a$, $\omega$ and $\lambda$ are
positive constants.  The degree of freedom $x$ remains completely classical in
the original treatment and serves as a background which, while rolling down
its linear potential $-ax$, excites the oscillator given by $z$ via the
interaction term $\frac{1}{2}\lambda x^2z^2$. In \cite{CQC}, $z$ is quantized
and its back-reaction effects on the roll-down rate of $x$ are studied.  This
dynamics could be considered a toy model representing the effects of particle
creation on a classical background.

Here, we extend the model in a way that goes beyond a strict background
treatment of $x$. In the cosmological situation, the usual separation into
background and perturbations exists only as an approximation of some
inhomogeneous dynamics. Because both background and perturbation variables
depend on the same fundamental degrees of freedom, given by the space-time
metric and suitable matter fields, a separate quantization has a certain
limited range of validity. Moreover, even classically the separation is not
compatible with full general covariance but is maintained only by coordinate
transformations that are small in the same sense in which inhomogeneity is
considered a small perturbation on the background. 

A complete treatment would have to be done within an elusive theory of quantum
gravity, but while such a theory is still being constructed, some generic
implications can be tested. In particular, if background and perturbations
depend on the same fundamental fields, upon quantization there may well be
correlations between them. Such correlations might even be required for
general covariance to hold at the quantum level: Since a small coordinate
change from $t$ to $t+\xi$ with an inhomogeneous $\xi\ll t$, allowed at the
perturbative level, maps a pure background variable such as the scale factor
$a(t)$ into an inhomogeneously perturbed quantity, $a(t+\xi)\approx
a(t)+\dot{a}\xi$, assuming vanishing quantum correlations between background
and perturbations is likely too restrictive for a covariant quantum theory of
the system.

The task then is to introduce correlation parameters between background and
perturbations while maintaining a suitable level of tractability. We solve
this problem here by utilizing a quasiclassical formulation of quantum
dynamics \cite{VariationalEffAc,EnvQuantumChaos,GaussianDyn,QHDTunneling}
which introduces (or derives) canonical variables for moments of a quantum
state. (As we will also demonstrate, the constructions of \cite{CQC} are
equivalent to these quasiclassical methods.) Recent advances in
\cite{Bosonize,EffPotRealize} have extended these methods from a single degree
of freedom to a pair of coupled and possibly correlated degrees of
freedom. This extension makes it possible to derive a systematic formulation
of cosmological perturbations with background correlations.

In \cite{CQCFields}, a quantum-field version of the model of \cite{CQC} was
introduced and studied, concluding that it was not possible to solve the
homogeneity problem. We describe and extend related models by applying
systematic quasiclassical methods to a quantum field on a semiclassical
background. Also here, we are able to introduce correlations between the field
and the background. Importantly, we will also be able to reveal a new effect
implied by the evolution of cross-correlations of different modes of a quantum
field which can be used to address the homogeneity problem.  In particular, we
will observe that the usual assumption of an exactly homogeneous initial state
is not always justified because super-Hubble inhomogeneity (as implied at
early times by the BKL scenario) may trickle down to within the Hubble radius
when quantum correlations are considered.  We interpret this new effect as a
consequence of quantum non-locality.

\section{Canonical description of quantum dynamics}

In order to facilitate the mathematical description of a quantum degree of
freedom interacting with a classical variable, the construction presented in
\cite{CQC} uses a semiclassical approximation. Motivated by the well-known
form of energy eigenstates of the harmonic oscillator, in particular of the
Gaussian ground state, the authors observe that the first semiclassical
correction of an oscillator can be described by doubling its degrees of
freedom. That is, instead of a single variable $z$, one may use a pair of
degrees of freedom, $\chi$ and $\xi$ with momenta $p_{\chi}$ and $p_{\xi}$,
such that the quadratic potential $z^2$ is replaced by $\chi^2+\xi^2$.  The
equations of motion are therefore given by
\begin{equation}\label{cqc-eom}
  \ddot{x}=a-x\lambda (\xi^2+\chi^2)\quad,\quad
  \ddot{\xi}=-(\omega^2+\lambda x^2)\xi\quad,\quad
  \ddot{\chi}=-(\omega^2+\lambda x^2)\chi \,.
\end{equation}

In addition, the angular momentum in the new $(\chi,\xi)$-plane is constrained
to be non-zero and equal to $\xi p_{\chi}- \chi p_{\xi}=1/2$, restricting the
allowed initial values. In particular, it is not possible for only one of the
new variables to be non-zero, in which case we would have equivalence with the
classical formulation.

In \cite{CQC}, this doubling of degrees of freedom of a semiclassical particle
is obtained by including the variance as an independent degree of freedom, in
addition to the expectation value of a basic variable such as $z$ (which
vanishes in an energy eigenstate). The expectation value of $\hat{p}_z^2$ in
an uncorrelated Gaussian state with variance $\sigma^2$ is given by
\begin{equation}
 \langle\hat{p}_z^2\rangle= \langle\hat{p}_z\rangle^2+ \frac{\hbar^2}{4\sigma^2}
\end{equation}
while the expectation value of $\hat{z}^2$ (in any state) equals
\begin{equation}
 \langle\hat{z}^2\rangle= \langle\hat{z}\rangle^2+\sigma^2\,.
\end{equation}
The Gaussian expectation value of (\ref{H}) in which only $(z,p_z)$ has been
quantized is therefore
\begin{equation} \label{H1}
	\langle\hat{H}\rangle=
        \frac{1}{2}p_x^2+\frac{1}{2}\langle\hat{p}_z\rangle^2+
        \frac{\hbar^2}{8\sigma^2}  -ax+\frac{1}{2}(\omega^2+\lambda  
        x^2)(\langle\hat{z}\rangle^2+ \sigma^2) \,. 
\end{equation}
The contribution $\hbar^2/(8\sigma^2)$ may be interpreted as a centrifugal
potential of planar motion expressed in polar coordinates with radius
$\sigma$, together with a spurious angle that does not appear in the potential
of (\ref{H1}).  Transforming to Cartesian coordinates implies the two degrees
of freedom, $\chi$ and $\xi$ such that $\chi^2+\xi^2=\sigma^2$, with the
condition that their angular momentum has to equal $\hbar/2$ for
$\hbar^2/(8\sigma^2)$ in (\ref{H1}) to be the correct centrifugal
potential. The harmonic potential is then turned into
$\frac{1}{2}(\omega^2+\lambda x^2)(\langle\hat{z}\rangle^2+ \xi^2+\chi^2)$,
which implies the equations of motion (\ref{cqc-eom}) for vanishing
expectation value $\langle\hat{z}\rangle$, as assumed in (\ref{cqc-eom}).  To
complete this construction, we also need a kinetic energy of the new variable
$\sigma$ or its Cartesian analogs $\xi$ and $\chi$, which will be provided
naturally by our more general treatment below.

For comparison, we briefly recall an alternative derivation of the equations
of motion closer to the approach of \cite{CQC}:
The two degrees of freedom can be derived directly by
introducing Heisenberg operators
\begin{eqnarray}
	\hat{x}(t)&=& z(t)^*\hat{a}_0+z(t)\hat{a}_0^{\dagger} \\
	\hat{p}(t)&=& \dot{z}(t)^*\hat{a}_0+\dot{z}(t)\hat{a}_0^{\dagger}
\end{eqnarray}
with the time-independent annihilation operator $\hat{a}_0$ and a
time-dependent complex function $z=\xi+i\chi$.
Commutation relations then imply that
\begin{eqnarray}
	\hat{x}^2&=& z^*z(1+2\hat{n}_0)+(z^*)^2\hat{a}_0^2+z^2(a_0^{\dagger})^2\\
	\hat{p}^2&=& \dot{z}^*\dot{z}(1+2\hat{n}_0)+(\dot{z}^*)^2\hat{a}_0^2
	+\dot{z}^2(\hat{a}_0^{\dagger})^2\\
\end{eqnarray}
where $\hat{n}_0$ is the time-independent number operator.
In the ground state, which we assume initially,
where $\langle\hat{x}\rangle=0=\langle\hat{p}_z\rangle$,
we have
\begin{eqnarray}
	\label{x-map}
	\Delta(x^2)\equiv \langle(\hat{x}-\langle \hat{x}\rangle)^2\rangle
	&=& \xi^2+\chi^2\\
	\label{p-map}
	\Delta(p^2)\equiv \langle(\hat{p}-\langle \hat{p}\rangle)^2\rangle
	&=& \dot{\xi}^2+\dot{\chi}^2\,,
\end{eqnarray}
valid at all times.
Fluctuations therefore imply two dynamical real degrees of freedom. Moreover,
the covariance (in the standard sense) equals
\begin{equation} \label{cross-term}
	\Delta(xp)=
        \frac{1}{2}\langle\hat{x}\hat{p}+\hat{p}\hat{x}\rangle=
        \xi\dot{\xi}+\chi\dot{\chi} 
\end{equation}
such that the uncertainty product
\begin{equation}
	\Delta (x^2)\Delta(p^2)-\Delta(xp)^2=(\xi\dot{\chi}-\dot{\xi}\chi)^2
\end{equation}
implies a constant $\xi\dot{\chi}-\dot{\xi}\chi= \hbar/2$ in the Gaussian
ground state. The specific angular momentum required by (\ref{H1}) is
therefore closely related to the uncertainty relation.

We point out that the canonical formulation underlying this procedure has been
known for some time, and has in fact been (re)discovered independently in
various fields, including quantum field theory \cite{VariationalEffAc},
quantum chaos \cite{EnvQuantumChaos} and quantum chemistry
\cite{GaussianDyn,QHDTunneling}.  Leading semiclassical corrections of a
particle moving in one dimension with coordinate $z$ and momentum $p_z$ can be
described by coupling the basic expectation values $\langle\hat{z}\rangle$ and
$\langle\hat{p}_z\rangle$ to three additional variables, the quantum
fuctuations $\Delta(z^2)$ and $\Delta(p_z^2)$, as well as the covariance
$\Delta(zp_z)=\frac{1}{2}\langle \hat{z}\hat{p}_z+\hat{p}_z\hat{z}\rangle-
\langle\hat{z}\rangle\langle\hat{p}_z\rangle$. It is useful to introduce a
uniform notation that can easily be extended to higher moments, which we
write, following \cite{EffAc,EffConsRel,Counting}, as
\begin{equation}
 \Delta(z^ap_z^b)= \langle (\hat{z}-\langle\hat{z}\rangle)^a
 (\hat{p}_z-\langle\hat{p}_z\rangle)^b\rangle_{\rm Weyl}
\end{equation}
in completely symmetric, or Weyl ordering.

According to \cite{EffAc,Karpacz}, the expectation values and moments form a
phase space equipped with a Poisson bracket defined by
\begin{equation} \label{Poisson}
 \{\langle\hat{A}\rangle,\langle\hat{B}\rangle\}=
 \frac{\langle[\hat{A},\hat{B}]\rangle}{i\hbar}
\end{equation}
and extended to moments by using the Leibniz rule. As a simple consequence,
the Poisson bracket of basic expectation values equals the classical Poisson
bracket, $\{\langle\hat{z}\rangle,\langle\hat{p}_z\rangle\}=1$, and moments
have zero Poisson brackets with basic expectation values. A closed-form
expression exists for the bracket of two moments \cite{EffAc,HigherMoments},
but it is rather complicated. In particular, it is not in canonical form.

For instance, for second-order moments, we have the brackets
\begin{equation} \label{Poisson2}
 \{\Delta(z^2),\Delta(zp_z)\}=2\Delta(z^2)\quad,\quad
 \{\Delta(zp_z),\Delta(p_z^2)\}= 2\Delta(p_z^2)\quad,\quad
 \{\Delta(z^2),\Delta(p_z^2)\}= 4\Delta(zp_z)\,.
\end{equation}
With hindsight, the semiclassical formulation of
\cite{VariationalEffAc,EnvQuantumChaos,GaussianDyn,QHDTunneling} can be
interpreted as a mapping from the 3-dimensional Poisson manifold with brackets
(\ref{Poisson2}) to canonical, or Casimir--Darboux coordinates. Explicitly,
defining the mapping from $(\Delta(z^2),\Delta(zp_z),\Delta(p_z^2))$ to
$(s,p_s,U)$ by
\begin{equation}
 s=\sqrt{\Delta(z^2)} \quad,\quad p_s=
 \frac{\Delta(zp_z)}{\sqrt{\Delta(z^2)}}\quad,\quad
   U=\Delta(z^2)\Delta(p_z^2)-\Delta(zp_z)^2
\end{equation}
or its inverse,
\begin{equation} \label{sps}
 \Delta(z^2)=s^2\quad,\quad \Delta(zp_z)=sp_s\quad,\quad
 \Delta(p_z^2)=p_s^2+\frac{U}{s^2}\,,
\end{equation}
one can see that we have the canonical Poisson bracket $\{s,p_s\}=1$, while
$\{s,U\}=\{p_s,U\}=0$. 

These equations hold for all states. If we make the additional assumption that
second-order moments provide a good approximation of quantum dynamics at least
for some time, we may insert (\ref{sps}) in the expectation value of the
harmonic Hamiltonian, taken in an arbitrary semiclassical state. We then
obtain the effective Hamiltonian
\begin{equation} \label{H2}
	H_{\rm eff}=\langle\hat{H}\rangle=
\frac{1}{2}p_x^2-ax+\frac{1}{2}p_z^2+\frac{1}{2}(\omega^2+\lambda
        x^2)z^2+ 
	\frac{1}{2}\left(p_s^2+\frac{U}{s^2}\right)+\frac{1}{2}(\omega^2+\lambda
        x^2)s^2 \,, 
\end{equation}
still quantizing only $(z,p_z)$. This Hamiltonian is equivalent to (\ref{H1})
if $U=\hbar^2/4$, the minimum value allowed by Heisenberg's uncertainty
relation. It is more general if $U$ is allowed to be greater than this value,
in which case we are no longer restricted to Gaussian states. The derivation
shows how the conserved quantity $U$ is related to the uncertainty relation as
well as angular momentum in an effective description after transforming to
Cartesian coordinates.
Tranforming $s$ as the radial coordinate in an auxiliary plane (together with
a spurious angle) to Cartesian coordinates $(\xi,\chi)$ on this plane, the
centrifugal potential $U/(2s^2)$ can be eliminated by doubling the fluctuation
degree of freedom $s$:
\begin{equation} \label{HCartesian}
	H_{\rm Cartesian}=
\frac{1}{2}p_x^2-ax+\frac{1}{2}\left(p_z^2+p_{\xi}^2+p_{\chi}^2\right)
+\frac{1}{2}(\omega^2+\lambda
        x^2)(z^2+\xi^2+\chi^2) \,. 
\end{equation}
The kinetic energy of $s$, or $\xi$ and $\chi$, is automatically provided by
(\ref{H2}).  In general, angular momentum is bounded from below but not
required to equal $\hbar/2$ for generic states.

The effective Hamiltonian generates equations of motion for $x$, $z$ and $s$,
as well as their momenta.  Semiclassical aspects of quantum evolution can
therefore be described by an enlarged phase space of classical type. Compared
with the classical equations, solutions require additional initial values
which partially encode properties of quantum states.  The specification of an
arbitrary state would require infinitely many parameters, for instance all
moments required for the Hamburger problem that asks how a probability density
can be reconstructed from all its moments. A semiclassical approximation
replaces this infinite number with finitely many values, given by a minimum of
three non-classical parameters $s$, $p_s$, and $U$ at leading semiclassical
order.

\begin{figure}	
	\centering
	\includegraphics[width=\textwidth]{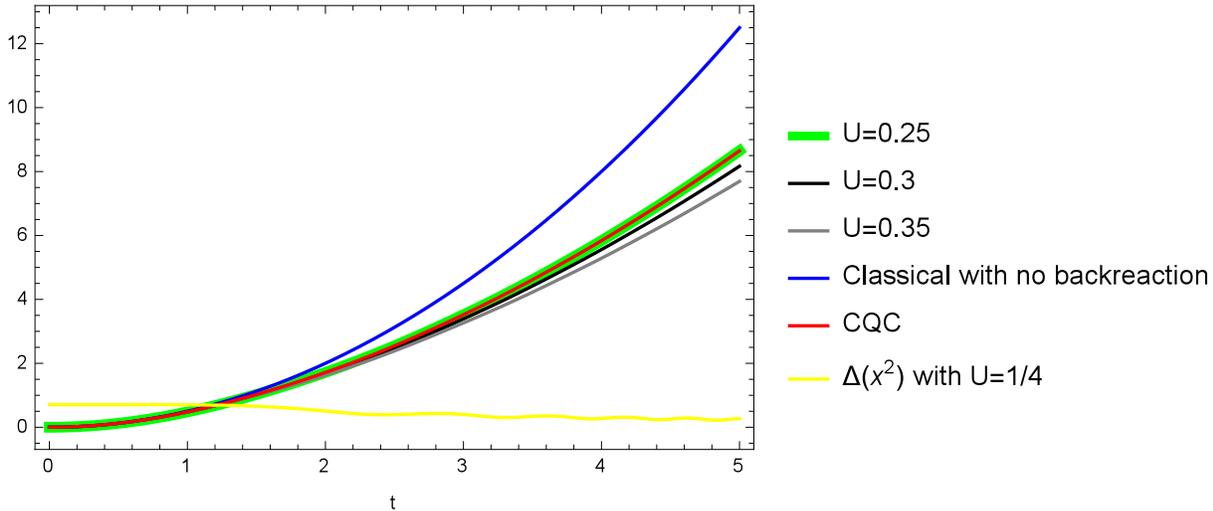}
	\caption{Classical evolution (blue) of the background variable $x(t)$,
          coupled to a semiclassical oscillator according to (\ref{H}),
          compared with semiclassical evolution generated by (\ref{H2}) for
          the values $U=0.25$, $0.3$ and $0.35$. Minimal uncertainty
          ($U=0.25$, thick
          green) agrees with the formulation of \cite{CQC} (red), while other
          values of $U$ lead to different dynamics. The yellow
          curve shows the evolving quantum fluctuation $s$ of the coupled
          oscillator for the case of $U=0.25$.  To fix units, the same choices
          $\omega=\lambda=\hbar=1$ as in \cite{CQC} have been made.}
\label{rolling-fig}
\end{figure}

A simple initial state, which may be Gaussian but would not be required to
stay so in an interacting system, can be specified by the choice
\begin{equation}\label{IV}
 p_x(0)=x(0)=0\quad,\quad p_z(0)=z(0)=0\quad,\quad
 s(0)=\frac{1}{\sqrt{2\omega}} \quad,\quad p_s(0)=0 \,.
\end{equation}
The specific value chosen for $s$ mimics the ground state of a harmonic
oscillator with frequency $\omega$.  We may leave the Casimir $U$ as a free
parameter, which is restriced by the inequality $U\geq \hbar^2/4$ but need not
saturate it if the state is not required to be Gaussian. It would then be more
difficult to find a specific wave function that belongs to these parameters,
but semiclassical evolution based on the equations given here can be performed
without problems. As shown in Fig.~\ref{rolling-fig}, the dynamics of the
model applied in \cite{CQC} to Gaussian states is indeed sensitive to the
value of $U$.

\section{Correlations with the background}

Having established the close relationship between \cite{CQC} and canonical
effective methods, we now use extensions of the latter to generalize the
dynamics by including correlation degrees of freedom.

\subsection{Motivation}

In a traditional background treatment, the variable $x$ is treated completely
classically, as in \cite{CQC} and reviewed in the preceding section. However,
if the system with Hamiltonian (\ref{H}) is used as a toy model for particle
production and backreaction in cosmology, the distinction between background
degrees of freedom and perturbations is not as clear-cut because the model is
then part of a generally covariant system subject to symmetries that act on
matter fields as well as the space-time metric. A clear separation between
background and perturbations as distinct degrees of freedom is usually
available only in a fixed gauge, but not in a physical or gauge-invariant
manner. For gauge-invariant properties, it is therefore of interest to go
beyond a pure background treatment.

In particular, one commonly uses curvature perturbations or Mukhanov--Sasaki
variables in order to put perturbative inhomogeneity into ``gauge-invariant''
form \cite{CosmoPert}. (In the present discussion, we may ignore the fact that
these expressions are not fully gauge-invariant, but are so only with respect
to a subset of all gauge transformations even in the linearized setting
\cite{Stewart,NonCovDressed}.) These variables combine scalar modes of the
metric with scalar fields. The former degrees of freedom form a single
tensorial object together with the background metric, and therefore pure
matter perturbations, distinct from the background, can be obtained only if a
gauge is chosen in which the scalar modes of the metric vanish. In a
gauge-invariant treatment, by contrast, it is not clear in which sense
background and perturbations may be considered sufficiently independent to
justify the assumption of vanishing correlations in a generic state upon
quantization. (Thanks to general covariance, the quantum description of
cosmological perturbations on an expanding background is not the same as
quantum field theory on a curved background space-time, in which form it is
often presented.)

In fact, the expressions for curvature perturbations in terms of metric and
matter fields depend on the background scale fcator and the Hubble
parameter. Background and perturbations are therefore not independent in a
framework that somehow derives quantized perturbations from some fundamental,
unperturbed quantum theory of gravity. Such a derivation would, of course, be
challenging, but it suggests that background correlations should be relevant:
Unless it can be shown that a quantum gravity could not possibly lead to
quantum correlations between background and perturbations, it is not justified
to assume that such correlations are absent or can be ignored. In this
section, we derive a new theory of canonical effective equations with
correlations, still applied to the toy Hamiltonian (\ref{H}).

\subsection{Canonical variables for second-order moments of two degrees of
  freedom} 

The technical task is then to generalize the simple mapping (\ref{sps}) to the
moments of two classical degrees of freedom, $x$ with momentum $p_x$ and $z$
with momentum $p_z$. To second order, each canonical pair has three individual
moments, given by two fluctuations and a position-momentum covariance. These
six individual moments are accompanied by four cross-covariances that contain
one variable from each pair, such as $\Delta(xz)$. We therefore have a
ten-dimensional Poisson manifold, which requires some work to put it into
canonical form with Darboux coordinates and Casimir variables. This derivation
has been completed only recently, in \cite{Bosonize,EffPotRealize}, where
explicit expressions for the ten second-order moments in terms of four pairs
of Darboux coordinates --- $(s_1,p_{s_1})$, $(s_2,p_{s_2})$,
$(\alpha,p_{\alpha})$ and $(\beta,p_{\beta})$ --- and two Casimir variables,
$C_1$ and $C_2$, have been provided.

The resulting expressions are rather complicated and, in contrast to
(\ref{sps}), have momentum variances that cannot be put into a form suitable
for canonical kinetic energies with constant coefficients. (A proof of this
claim has been provided in \cite{Bosonize}.) Fortunately, as we will show
here, they can be reduced in an approximate way that provides canonical
kinetic energies and retains a single independent correlation parameter,
$\beta$.

We will quote here only the relevant moments. The new fluctuation parameters
$s_1$ and $s_2$ as well as the correlation parameter $\beta$ are introduced by
the equations
\begin{equation}  \label{ssbeta}
 \Delta(x^2)=s_1^2 \quad,\quad \Delta (z^2)=s_2^2 \quad,\quad \Delta(xz)=
 s_1s_2\cos\beta\,, 
\end{equation}
straightforwardly generalizing (\ref{sps}). Momentum variances also have a
form similar to (\ref{sps}), given by 
\begin{equation} 
 \Delta(p_z^2)=p_{s_2}^2+ \frac{U_2}{s_2^2}
\end{equation}
for the variance of $p_z$. In contrast to (\ref{sps}), however, $U_2$ is not
constant  but rather depends on the canonical (Darboux) coordinates in the
complicated form
\begin{equation} \label{U2}
 U_2=(p_{\alpha}-p_{\beta})^2+\frac{1}{2 \sin^2{\beta}}\left((C_1-4
  p_{\alpha}^2)-\sqrt{C_2-C_1^2+(C_1-4 p_{\alpha}^2)^2}\sin{(\alpha+\beta)}\right)
\end{equation}
where $C_1$ and $C_2$ are the two Casimir variables, and therefore
conserved. A similar expression exists for $\Delta(p_x^2)$.

This rather long expression has been derived in \cite{EffPotRealize} from the
conditions that the moments, expressed in canonical variables, (i) obey the
required Poisson brackets and (ii) are represented in a one-to-one manner
without loosing degrees of freedom in the canonical parameterization. This
process, in particular the second condition, requires the inclusion of a
canonical pair $(\alpha,p_{\alpha})$ in (\ref{U2}) whose physical
interpretation is not as clear as that of the fluctuation and correlation
parameters $s_1$, $s_2$ and $\beta$ together with their momenta. 

A possible interpretation can be obtained from the fact that moments can be
defined for any pure or mixed state. While the known meaning of $s_1$, $s_2$
and $\beta$ as well as their momenta shows that they are free parameters even
if one restricts oneself to pure states, there must be additional parameters
in a parameterization of all states that determine how much they deviate from
a pure state. The new canonical coordinates $\alpha$ and $p_{\alpha}$, as well
as $C_2$, are candidates for such impurity parameters, a conjecture which has
been tested with some success in a new canonical derivation of low-energy
effective potentials given in \cite{EffPotRealize}.

\subsection{Reduction of degrees of freedom}

A minimal model that goes beyond the mapping for a single degree of freedom
and retains a correlation parameter can be constructed as follows: We first
assume that $p_{\alpha}$ and $\sqrt[4]{C_2}$ are much smaller than $p_{\beta}$
and $\sqrt{C_1}$. The square root in (\ref{U2}) is then small, suppressing the
dependence on $\alpha$ via $\sin(\alpha+\beta)$.  The variable $\alpha$
therefore need not be assumed small, and it may in fact grow because,
according to the cross-term of the first square in $U_2$, any effective
Hamiltonian to which $\Delta(p_z^2)$ contributes in the kinetic energy,
generates an equation of motion of the form $\dot{\alpha}\propto
p_{\beta}+\cdots$ where dots indicate terms independent of $p_{\beta}$. It is
therefore impossible for $\alpha$ to be exactly zero if $p_{\beta}$ (a
correlation parameter like $\beta$) is non-zero, while we need a non-zero
$p_{\beta}$ in order to consistently ignore $p_{\alpha}$ unless this variable
is exactly zero. However, $\alpha$ appears in a bounded function in (\ref{U2})
that is suppressed by a small square root if our assumption about $p_{\alpha}$
and $C_2$ is satisfied, such that the dependence on $\alpha$ can be ignored in
this case.  The interpretation of $\alpha$, $p_{\alpha}$ and $C_2$ as impurity
parameters suggests that our approximation should be valid whenever a state is
close to being pure.

We then have the simplified expression
\begin{equation}  \label{Deltap2}
 \Delta (p_2^2)= p_{s_2}^2+ \frac{p_{\beta}^2}{s_2^2}+
 \frac{C_1}{2s_2^2\sin^2\beta} 
\end{equation}
which, in a generalization of the momentum terms in (\ref{H2}), can be
interpreted as (twice) the kinetic energy of a particle moving in three
dimensions, expressed in spherical coordinates $(s_2,\beta,\varphi)$ with a
spurious degree of freedom $\varphi$. In contrast to the Cartesian version of
(\ref{H2}), one of the angles, $\beta$, now is physically meaningful. It is,
in fact, the correlation parameter relevant for our present aims. 

As shown by the last term in (\ref{Deltap2}), the momentum $p_{\varphi}$ of
the spurious angle is constrained to equal the constant $\sqrt{C_1/2}$. If we
include the second degree of freedom in the mapping, given by the background
variable $x$, it would have a similar kinetic energy with the same angle
$\beta$ and its momentum $p_{\beta}$. The two 3-dimensional systems would
therefore be subject to constraints.

Here, we apply the 3-dimensional model only to the oscillating degree of
freedom, $z$, while the background degree of freedom $x$ is extended only by a
fluctuation variable, $s$, as in a mapping for a single degree of freedom. In
this way, we are able to construct a minimal extension of the model by
parameterizing the correlation variable, $\beta$. For $x$, we therefore have
the kinetic contribution
\begin{equation}
 \frac{1}{2}\left(p_x^2+p_s^2+ \frac{U}{s^2}\right)=
 \frac{1}{2}(p_x^2+p_X^2+p_Y^2) 
\end{equation}
as before, implying the contribution
\begin{equation}
 -ax+ \frac{1}{2} \lambda s^2z^2= -ax+\frac{1}{2} \lambda (X^2+Y^2)z^2
\end{equation}
to the effective potential in Cartesian coordinates, in the style of
\cite{CQC} but now applied also to the background. 

For $z$, we transform the kinetic energy implied by (\ref{Deltap2}) to
3-dimensional Cartesian coordinates $(\xi,\chi,\zeta)$, such that
\begin{equation}
 \frac{1}{2}\left(p_z^2+p_{s_z}^2+ \frac{U_2}{s_z^2}\right)= \frac{1}{2}
 (p_z^2+p_{\xi}^2+p_{\chi}^2+p_{\zeta}^2) \,.
\end{equation}
This variable contributes several terms to the effective potential:
\begin{equation}
 \frac{1}{2} (\omega^2+\lambda x^2)(z^2+\xi^2+\chi^2+\zeta^2)+
 2\lambda xzs\zeta 
\end{equation}
where the last term comes from the correlation (\ref{ssbeta}).
The Hamiltonian is therefore
\begin{eqnarray} \label{HCorr}
 H&=& \frac{1}{2}(p_x^2+p_z^2+p_{X}^2+p_{Y}^2+p_{\xi}^2+p_{\chi}^2+p_{\zeta}^2) -ax
 + \frac{1}{2} (\omega^2+\lambda
 x^2)(z^2+\xi^2+\chi^2+\zeta^2)\nonumber\\
&&+ \frac{1}{2}\lambda
 \left(z^2(X^2+Y^2)+4xz\sqrt{X^2+Y^2} \zeta\right)\,.
\end{eqnarray}
For zero cross-correlations, we have $\zeta=0$ and the Hamiltonian is reduced
to a strict background model if we also set $X^2+Y^2=0=p_X^2+p_Y^2$.

\subsection{Diagonalization}

The new effective Hamiltonian (\ref{HCorr}) can be interpreted as a system of
four harmonic oscillators --- $z$, $\xi$, $\chi$ and $\zeta$ --- with
frequencies that depend on time through the background variable $x$ and its
fluctuation parameter, $\sqrt{X^2+Y^2}$.  The term $xz\sqrt{X^2+Y^2}\zeta$ in
(\ref{HCorr}) implies that $z$ and $\zeta$ are not normal coordinates of the
oscillator system. 

We may attempt to (time-dependently) diagonalize this coupling between $z$ and
the component $\zeta$ of its fluctuation/correlation. In order for the
diagonalization to be feasible, we assume that the typical time scale of
evolution for $z$ and $\zeta$ is much smaller than the time scale of $x$ and
its fluctuations. We will then be able to treat the coefficients as
approximately time-independent, allowing a straightforward diagonalization of
the quadratic form.  

Considering the entire Hamiltonian $H$, the time scale for $x$ is of the order
$1/a\sim O(1)$ (noting that we set the mass to one and use natural units),
while the time scale for $\Delta(x^2)=X^2+Y^2$ is of the order $1/(\lambda
z^2)^{1/2}\gg 1$, as will be confirmed in Fig. \ref{z1z22}. Turning to $z$ and
$\zeta$, we see that their time scales are roughly $1/(\lambda x^2)^{1/2}$ and
$1/(x\sqrt{X^2+Y^2})^{1/2}$ which are typically very small because $x^2$ and
$\Delta(x^2)$ grow large at late times. The hierarchy in time scales, along
with the assumption $\omega^2\sim O(1)$, such that $\omega^2\ll\lambda x^2$,
justifies the following approximation. 

We may rewrite the $z-\zeta$ part of (\ref{HCorr}) as
\begin{equation}
	H_{z-\zeta}= K+\frac{1}{2}\lambda
 \left( (x^2+\delta^2) z^2+x^2\zeta^2\right)+
	2\lambda x\delta z\zeta
\end{equation}
where 
\begin{equation}
	\delta=\sqrt{X^2+Y^2}\quad\mbox{and}\quad
	K=\frac{1}{2}(p_z^2+p_{\zeta}^2)\,.
\end{equation}
Upon diagonalization we obtain the normal (angular) frequencies
\begin{eqnarray}
	\omega_1^2&=&\lambda \left(x^2+\frac{1}{2}\delta^2
        -\delta\sqrt{16x^2+\delta^2}\right)\\ 
	\omega_2^2&=&\lambda\left(x^2+\frac{1}{2}\delta^2+
    \delta\sqrt{16x^2+\delta^2}\right)\,. 
\end{eqnarray}
Since $\delta$ is the fluctuation of the classical degree of freedom, $x$, we
expect it to be much smaller than $x$ in magnitude. This result implies that
the typical behavior of evolution of $z$ and $\zeta$ can be described a fast,
nearly harmonic oscillation with frequency $\omega_1+\omega_2\approx
\sqrt{\lambda}x$, modulated by a slow oscillation with frequency
$\omega_2-\omega_1\approx 4\sqrt{\lambda}\delta$. The introduction of the
correlation parameter $\beta$ therefore gives rise to beat-like behavior,
which is new and only present if we capture the effect of the background
degree of freedom and its fluctuation using the 2-particle mapping.

\begin{figure}[!htb]
	\centering	
	\includegraphics[width=\textwidth]{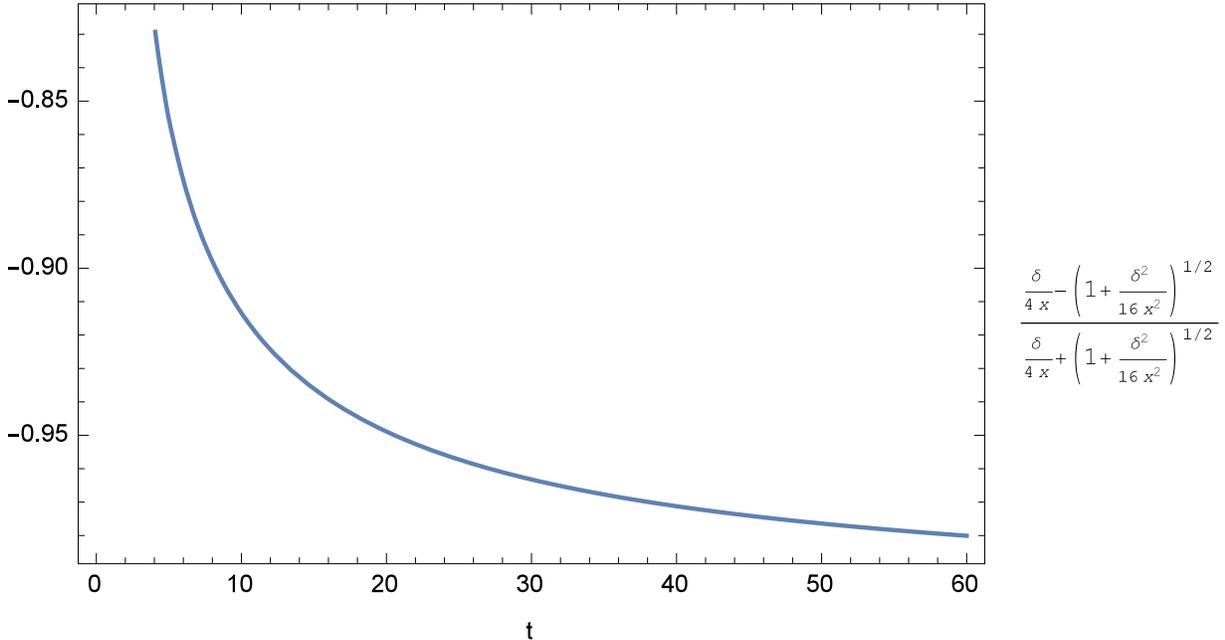}
	\caption{Behavior of the normal coordinate $e_1$ in (\ref{e1}).  The
          general trend indicates that $e_1\rightarrow -z+\zeta$
          asymptotically.  The parameters and initial values used here are
          specified in Section~\ref{s:DynImpl}.}
	\label{eigen1}
\end{figure}

The ``normal coordinates'' are given by
\begin{eqnarray}
	e_1&=& N_1\left(
          \left(\frac{\delta}{4x}-\sqrt{1+\frac{\delta^2}{16x^2}}\right)z 
	+\left(\frac{\delta}{4x}+\sqrt{1+\frac{\delta^2}{16x^2}}\right)
\zeta\right) \label{e1}\\
	e_2&=& N_2(z+\zeta)\,,
\end{eqnarray}
where $N_1$ and $N_2$ are normalization constants. We see in Fig. \ref{eigen1}
that the coefficients of $z$ and $\zeta$ in the first line grow to be of
similar magnitude but opposite signs, as a consequence of a decreasing
$\delta/x$ such that the background degree of freedom is becoming more and
more classical.

\subsection{Dynamical implications}
\label{s:DynImpl}

At the beginning of this section we showed that the semiclassical equations of
motion for agree with those of \cite{CQC}, but only if $z(0)=p_z(0)=0$ as
appropriate for a system in an initial vacuum state and only if a single
degree of freedom, $z$, is quantized.  We can mimic the same initial conditon
for the case where both degrees of freedom are quantized if we also impose
$\zeta(0)=p_{\zeta}(0)=0$, which we call the {\it equivalence conditions}.  We
also choose $Y(0)=\chi(0)=\frac{1}{\sqrt{2}}=p_X(0)=p_{\xi}(0)$ and other
variables initially $0$. The last two equalities for momenta are required by
the interpretation of the Casimir variables identified with angular
momentum.

\begin{figure}
	\centering	
	\includegraphics[width=0.9\textwidth]{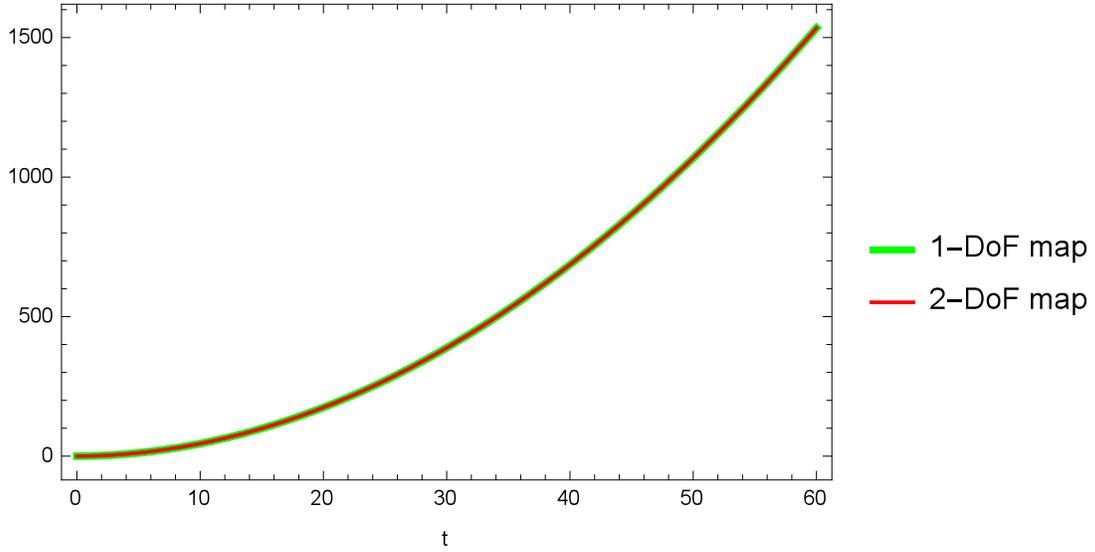}
	\caption{Numerical evolution for $\langle x(t)\rangle$ with (red) and without (green)
          background correlations, respectively, and equivalence conditions
          imposed. Starting with the 
          same initial values, the evolutions do not deviate from each other.
          \label{x1x2-1dof}}
\end{figure}

\begin{figure}
	\centering	
	\includegraphics[width=0.9\textwidth]{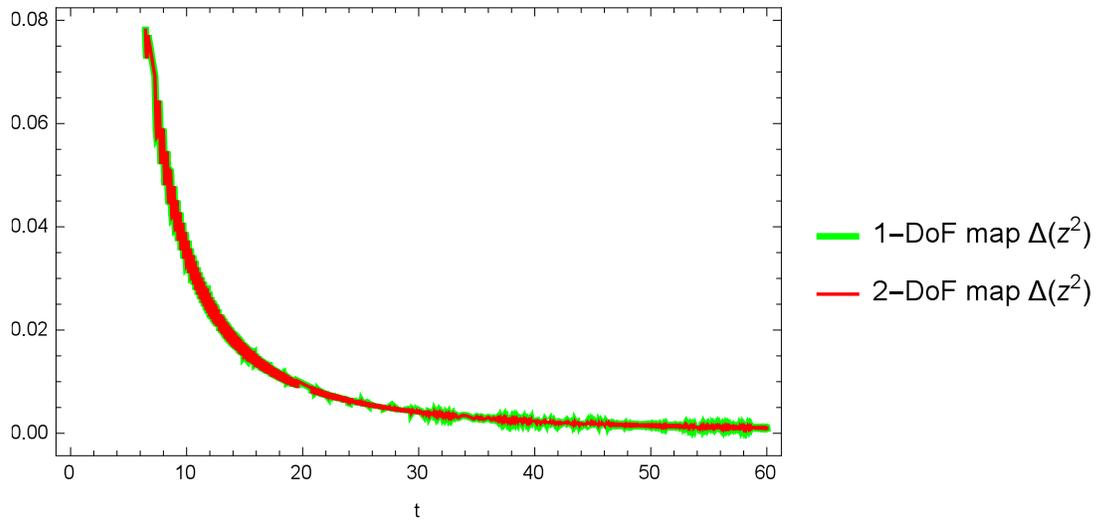}
	\caption{Fluctuations $\Delta(z^2)$ with (red) and without (green)
          background correlations, respectively, imposing equivalence
          conditions.}
	\label{deltaz-1dof}
\end{figure}

Numerical similations with these conditions do not reveal any additional
features as seen in Figs.~\ref{x1x2-1dof} and \ref{deltaz-1dof}.  Therefore,
the equivalence conditions turn the mapping for two degrees of freedom into a
system equivalent with \cite{CQC}. Even though we do introduce couplings of
$X,Y$ and $\zeta$ to $x$, the initial conditions $z(0)=0$ and hence
$\dot{p}_{\zeta}(0)=0$ imply $\zeta(t)=0$ through out evolution,
meaning that $x$ effectively
couples only to $\xi^2+\chi^2$, exactly as in the mapping for a single degree
of freedom.

\begin{figure}
	\centering	
	\includegraphics[width=0.9\textwidth]{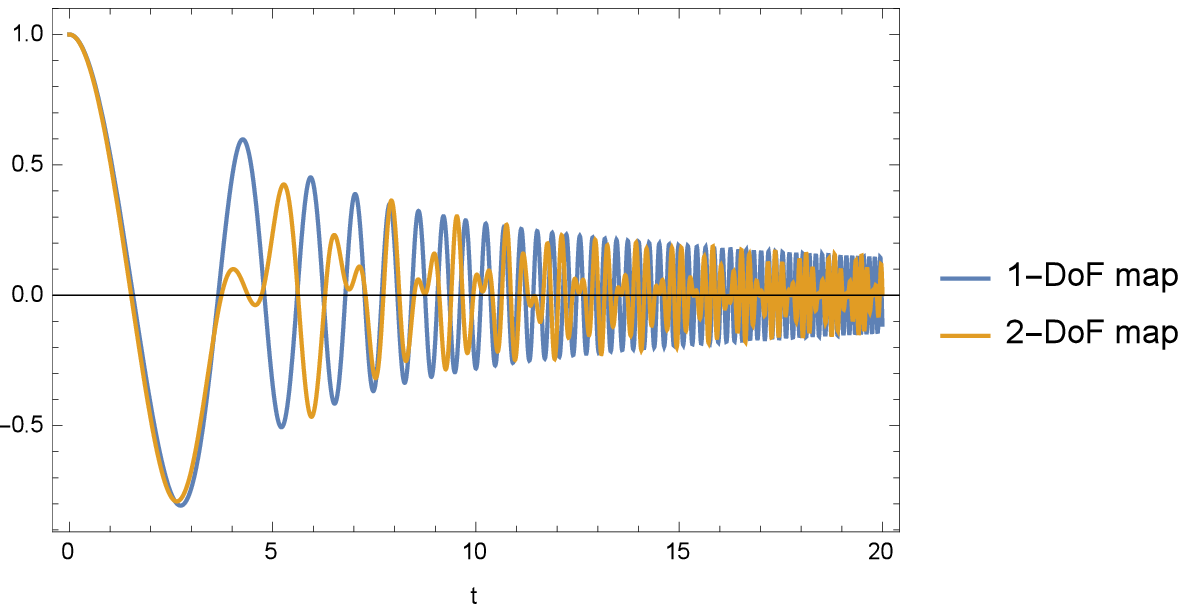}
	\caption{Expectation value $\langle z (t)\rangle $with (blue) and
          without (orange) background correlations, respectively.  Using the
          more general mapping for two degrees of freedom, we see a slow
          frequency modulation of the original the fast oscillations.  This is
          the beat-like behavior mentioned in the text.}
	\label{z1z22}
\end{figure}

\begin{figure}
	\centering	
	\includegraphics[width=0.9\textwidth]{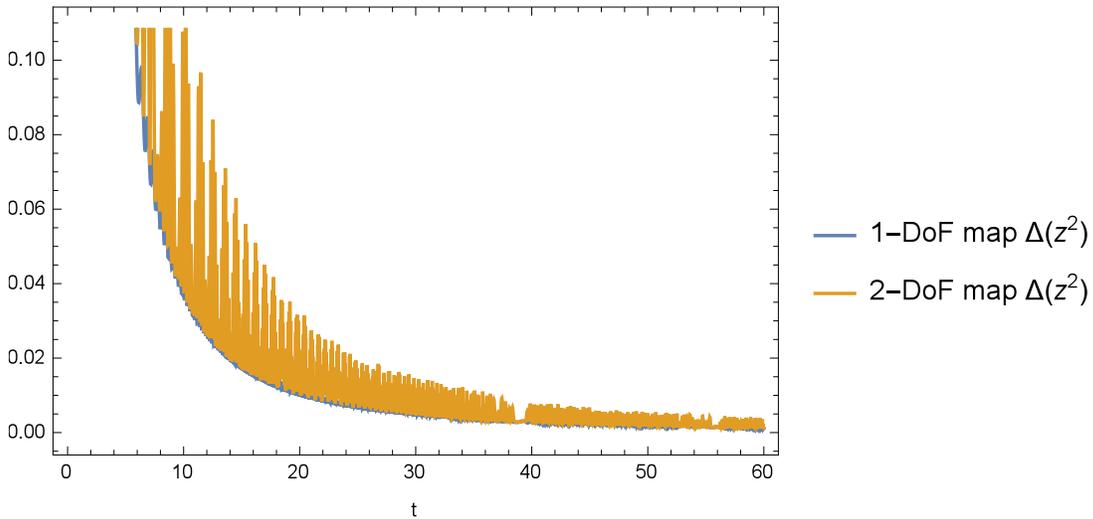}
	\caption{Fluctuation $\Delta(z^2)$ with (blue) and without (orange)
          background correlations, respectively.  There is again enhanced
          oscillation behavior due an additional dimension in fluctuation
          space given by the $\zeta$ direction.}
	\label{deltaz2}
\end{figure}

Interesting new effects are, however, obtained if the initial value of $z$ is
not zero, such that we do not start in a vacuum state. Physically, it could be
justified to use such initial values after a phase transition, where the
inflaton field acquires a non-zero vacuum expectation value.  For instance,
using $z(0)=1$, Figs.~\ref{z1z22} and \ref{deltaz2} show that both $\langle z
\rangle$ and $\Delta(z^2)$ obtain new oscillatory features from the mapping
for two degrees of freedom.  In addition to fast oscillations, $z(t)$ is also
modulated by a slow-frequency oscillation. The $x$-fluctuation, $\Delta(x^2)$,
also increases with time. Using (\ref{HCorr}), we can think of $\Delta(x^2)$
as a particle in a central-force problem, in which $\Delta(x^2)$ is subject to
a central force that decreases with time due to the decrease of $z^2$ and
$xz\zeta$. Since $\Delta(x^2)$ started off with a non-zero initial momentum
$p_X^2+p_Y^2$ due to the uncertainty relation, the particle (fluctuation) will
eventually escape to a larger distance from the center.  

\begin{figure}
	\centering	
	\includegraphics[width=0.9\textwidth]{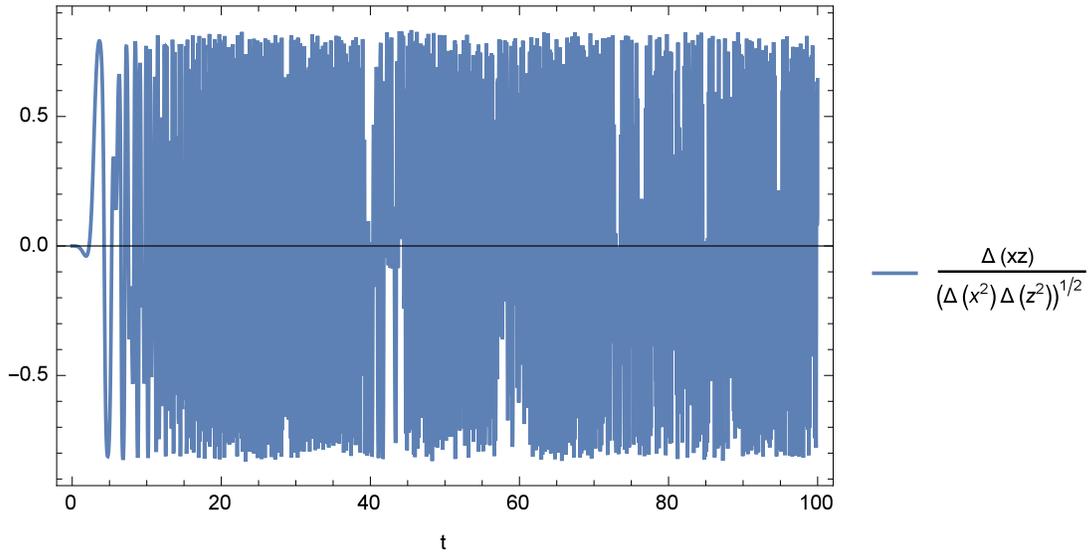}
	\caption{Time-dependent background correlation
          $\rho_{x,z}=\Delta(xz)/\sqrt{\Delta(x^2)\Delta(z^2)}$.  Its local
          maxima are near but not equal to one at late times. There is
          therefore a maximum correlation for the state implied by our
          initial conditions.}
	\label{corr}
\end{figure}

\begin{figure}
	\centering	
	\includegraphics[width=0.9\textwidth]{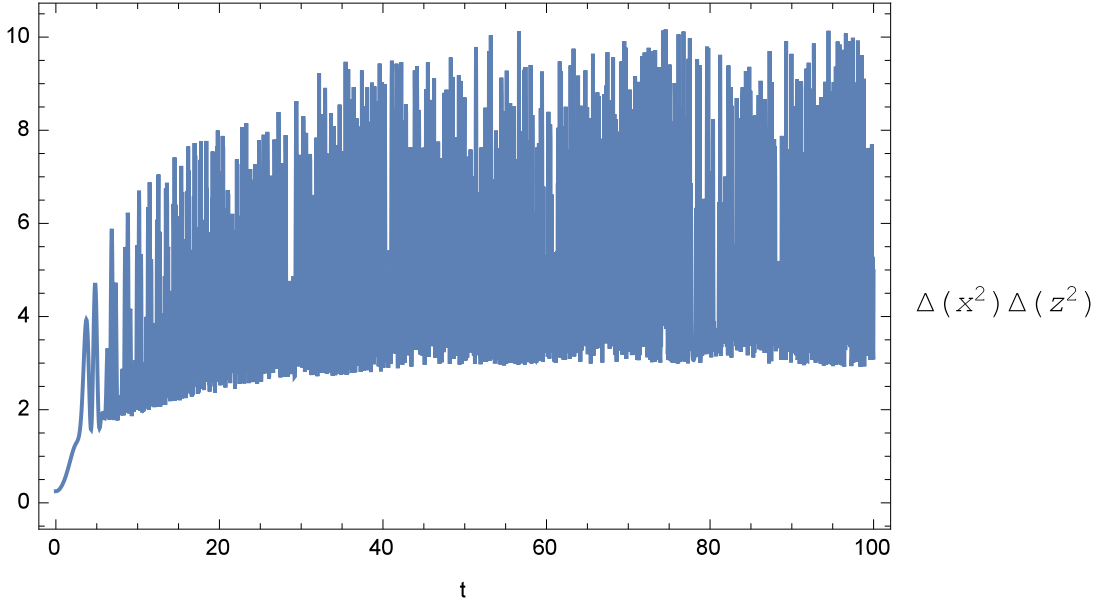}
	\caption{Boundedness of $\Delta(x^2)\Delta(z^2)$. Since $\Delta(x^2)$
          increases while $\Delta(z^2)$ decreases, the balance between the two
          in their product is unexpected.}
	\label{flucfluc}
\end{figure}

The correlation of the two degrees of freedom is shown in Fig.~\ref{corr}. Two
interesting features are the boundedness of $\Delta(x^2)\Delta(z^2)$ in
Fig.~\ref{flucfluc}, and the upper bound of the background correlation between
$z$ and $x$ for a given state in Fig.~\ref{corr}. The latter is explained by
the interpretation of the correlation parameter as a spherical angle in an
auxiliary space, implied by the appearance of (\ref{Deltap2}) in the form of
kinetic energy in spherical coordinates. The angular momentum $\sqrt{C_1/2}$
in this auxiliary space is concerved and generically non-zero for a given
initial state. For a non-zero value it is then impossible for the correlation
angle to get arbitrarily close to the poles of the spherical system, such that
$\cos\beta$ keeps a certain distance from its general limiting values
$\pm1$. This function equals the combination of moments plotted in
Fig.~\ref{corr}, confirming the reduced upper bound for the given state
implied by our initial values.

\section{Field theory model}

The formalism of \cite{CQC} has also been applied to a quantum field
back-reacting on a classical homogeneous background \cite{CQCFieldsHom}. In
this section, we will show how these methods are related to moments of a
quantum field; see \cite{CW} for a general formulation of quantum fields by
moments with a derivation of the Coleman--Weinberg potential
\cite{ColemanWeinberg}. The Coleman--Weinberg potential is conceptually
related to the setting considered in \cite{CQCFieldsHom} as it results from an
expansion of a quantum field around a homogeneous background expectation
value. The back-reaction equations of \cite{CQCFieldsHom} can therefore be
embedded in a canonical effective theory by a suitable extension of
\cite{CW}. In particular, the equations correspond to a leading-order
formulation of moment equations by canonical Darboux coordinates. In contrast
to the previous section, however, a complete Darboux formulation of a quantum
field is challenging because a single quantum field implies a multitude
independent degrees of freedom, which are hard to describe by canonical
variables for moments if all possible cross-correlations are
included. Nevertheless, an embedding of \cite{CQC} is feasible.

\subsection{Modes on a compact homogeneous background}

We consider a two-field model with a single spatial dimension, which will be
reduced to a quantum field $\psi$ and a classical field $\phi$. Both fields
are scalar and real. Extending the interactions of (\ref{H}), we introduce the
classical action 
\begin{equation} \label{S}
	S=\int {\rm d}t{\rm d}x
        \left(\frac{1}{2}\left(\dot{\phi}^2-(\partial_x\phi)^2\right) 
	-V(\phi)+\frac{1}{2}\left(\dot{\psi}^2-(\partial_x\psi)^2\right)
-\frac{1}{2}(m^2+\lambda\phi^2)\psi^2\right) 
\end{equation}
with the mass $m$ of $\psi$ and a coupling constant $\lambda$ that may be
interpreted as providing a $\phi$-dependent correction to the mass of
$\psi$. The field $\phi$ moves in a potential $V(\phi)$ which, generically,
may also include a mass term. As an extension of the preceding section,
however, we will continue to assume that $V(\phi)$ is linear in our numerical
examples.

In the quantum-mechanical model of the previous section,
back-reaction of $z$ on $x$ causes an exchange of energy: the quantum degree
of freedom $z$ back-reacts on the classical one and saps its energy, causing
$x$ to roll down more slowly on its linear potential. We expect similar
transfer in energy in the field version, where the energy lost by the
classical field excites particle production for the quantum field.

Using Legendre transformation we derive the Hamiltonian density
\begin{equation} \label{Hdens}
	\mathcal{H}=\frac{1}{2}\left(\Pi_{\phi}^2+(\partial_x\phi)^2\right)
+V(\phi) 
	+\frac{1}{2}\left(\Pi_{\psi}^2+(\partial_x\psi)^2+
\Omega_{\phi}(t,x)^2\psi^2\right) \,,
\end{equation}
introducing 
\begin{equation} \label{Omega}
 \Omega_{\phi}(t,x)=m^2+\lambda\phi^2(t,x)\,.
\end{equation}
In order to facilitate an analysis of particle production, we should expand
$\psi$ in Fourier modes with respect to a 1-dimensional, spatial wave number
$k$. The resulting system can then be interpreted as a background field,
$\phi$, coupled to a large number of oscillators with $\phi$-dependent mass
and frequency.

Fourier transforms of the basic canonical field are given by
\begin{eqnarray}
	\psi(x)&=&\frac{1}{\sqrt{2\pi}}\int^{\infty}_{-\infty}{\rm d}k
        e^{ikx}\tilde{\psi}(k)\\ 
	\Pi_{\psi}(x)&=&\frac{1}{\sqrt{2\pi}}
	\int^{\infty}_{-\infty}{\rm d}k e^{-ikx}\tilde{\Pi}_{\psi}(k)\,,
\end{eqnarray}
making time dependence implicit in this notation. The modes obey the reality
conditions
\begin{equation}
 \tilde{\psi}(-k) = \tilde{\psi}(k)^* \quad\mbox{and}\quad
 \tilde{\Pi}_{\psi}(-k)=\tilde{\Pi}_{\psi}(k)^*\,. 
\end{equation}
Choosing opposite signs in
the exponentials used to transform $\psi$ and $\Pi_{\psi}$, respectively,
simplifies the canonical structure of modes. In particular, the calculation
\begin{equation} \label{Poissonk}
 \int{\rm d}x \dot{\psi}\Pi_{\psi}= \frac{1}{2\pi} \int{\rm d}x
 \int_{-\infty}^{\infty} {\rm d}k \int_{-\infty}^{\infty} {\rm d}l e^{i(k-l)x}
 \dot{\tilde{\psi}}(k) \tilde{\Pi}_{\psi}(l)= \int_{-\infty}^{\infty} {\rm d}k
 \dot{\tilde{\psi}}(k) 
 \tilde{\Pi}_{\psi}(k) 
\end{equation}
implies that $\tilde{\Pi}_{\psi}(k)$ is canonically conjugate to
$\tilde{\psi}(k)$.

Keeping the $\phi$-Hamiltonian density $\mathcal{H}_{\phi}$ unchanged, the
Hamiltonian for Fourier modes is then given by
\begin{eqnarray}
	H&=&\int {\rm d}x \mathcal{H}_{\phi}
	+\frac{1}{4\pi}\int{\rm d}x{\rm d}k{\rm d}l 
	\left(e^{-i(k+l)x}\tilde{\Pi}_{\psi}(k)\tilde{\Pi}_{\psi}(l)
	+(-kl+\Omega_{\phi}(t,x)^2)e^{i(k+l)x}\tilde{\psi}(k)
\tilde{\psi}(l)\right)\nonumber\\ 
 &=& \int{\rm d}x \mathcal{H}_{\phi}
	+\frac{1}{2}\int_{-\infty}^{\infty}{\rm d}k
	\left(|\tilde{\Pi}_{\psi}(k)|^2+
	k^2|\tilde{\psi}(k)|^2+ N_k[\psi]\right)
\end{eqnarray}
with the non-local (in $k$-space) contribution
\begin{equation} \label{N}
	N_k[\psi]= \tilde{\psi}(k)\frac{1}{2\pi}
	\int_{-\infty}^{\infty}{\rm d}l  \int{\rm d}x
 \Omega_{\phi}(t,x)^2 e^{i(l+k)x}\tilde{\psi}(l)\,. 
\end{equation} 
In order to decouple different $k$ and obtain a local $k$-Hamiltonian, we now
assume that $\Omega_{\phi}(t,x)$ is homogeneous in $x$, such that the
$x$-integration in $N_k[\psi]$ results in a delta function that removes
non-locality. Since $\Omega_{\phi}(t,x)$ depends on the background field
$\phi(x)$ according to (\ref{Omega}), the background field is assumed to be
spatially homogeneous from now on. It may, however, be time-dependent. With
this assumption, the classical Hamiltonian
\begin{equation}
H=\int{\rm d}x \mathcal{H}_{\phi}
	+\frac{1}{2}\int_{-\infty}^{\infty}{\rm d}k
	\left(|\tilde{\Pi}_{\psi}(k)|^2+
	(k^2+\Omega_{\phi}(t)^2)|\tilde{\psi}(k)|^2\right)
\end{equation}
is local.

A final transformation introduces real fields in $k$-space by splitting
$\tilde{\psi}(k)$ and $\tilde{\Pi}_{\psi}(k)$ into real and imaginary parts,
\begin{eqnarray}
	\tilde{\psi}(k)&=& \frac{1}{\sqrt{2}}(A(k)+iB(k))\\
	\tilde{\Pi}_{\psi}(k)&=&\frac{1}{\sqrt{2}}(C(k)-iD(k))\,.
\end{eqnarray}
Reality conditions imply that $A(k)$ and $C(k)$ are even functions while
$B(k)$ and $D(k)$ are odd. 
Continuing the calculation in (\ref{Poissonk}), we have
\begin{eqnarray}
 \int_{-\infty}^{\infty}{\rm d}k \dot{\tilde{\psi}}(k)\tilde{\Pi}_{\psi}(k)&=&
 \frac{1}{2} 
 \int_{-\infty}^{\infty}{\rm d}k (\dot{A}(k)+i\dot{B}(k))
 (C(k)-iD(k))\\
 &=& \frac{1}{2}
 \int_{-\infty}^{\infty}{\rm d}k \left(\dot{A}(k)C(k)+\dot{B}(k)D(k)-i
 (\dot{A}(k)D(k)- \dot{B}(k)C(k))\right)\,.\nonumber
\end{eqnarray}
The imaginary contribution vanishes because it integrates an odd function
$\dot{A}(k)D(k)- \dot{B}(k)C(k)$ over the full real range of $k$. The real
contribution is even and can therefore be restricted to only positive $k$,
such that
\begin{equation}
 \int_{-\infty}^{\infty}{\rm d}k \dot{\tilde{\psi}}(k)\tilde{\Pi}_{\psi}(k)=
 \int_{0}^{\infty}{\rm d}k \left(\dot{A}(k)C(k)+\dot{B}(k)D(k)\right)\,.
\end{equation}
If we restrict to positive $k$, we therefore have the momenta $\Pi_A(k)=C(k)$
and $\Pi_B(k)=D(k)$ of $A(k)$ and $B(k)$, respectively.

Our final expression for the classial Hamiltonian in terms of real modes is
therefore
\begin{equation}
 H=\int {\rm d}x \mathcal{H}_{\phi}
	+\frac{1}{2}\int_0^{\infty} {\rm d}k \left(\Pi_{A}(k)^2+\Pi_{B}(k)^2
+\omega_{\phi}^2(k)(A(k)^2+B(k)^2)\right) \,,
\end{equation}
providing two harmonic oscillators per mode $k$, each with a time-dependent
frequency
\begin{equation}
 \omega_{\phi}(k)=\sqrt{k^2+\Omega_{\phi}(t)^2}
\end{equation}
that depends parameterically on the homogeneous background field $\phi$.  The
$\psi$-contribution to the Hamiltonian can directly be quantized to
\begin{equation} \label{Hphi}
	\hat{H}=\int{\rm d}x \mathcal{H}_{\phi}
	+\frac{1}{2}\int_{0}^{\infty}{\rm d}k
	\left(\hat{\Pi}_A(k)^2+\hat{\Pi}_B(k)^2+
	\omega_{\phi}(k)^2 (\hat{A}(k)^2+\hat{B}(k)^2) \right)\,.
\end{equation}

Numerical simulations of our dynamics will be simpler if we replace the
continuum of modes obtained so far with a discrete set by introducing periodic
boundary conditions in space. We therefore assume that space is compactified
to a circle with circumference $L$. This finite size also makes the
$\phi$-Hamiltonian well-defined for a homogeneous $\phi$. (The
$\phi$-Hamiltonian is described by a minisuperspace treatment, in which the
averaging size $L$ would play an important role in quantum corrections if
$\phi$ were quantized; see \cite{MiniSup,Infrared}. Here, it is relevant for a
consistent technical implementation of the background.) Our values of $k$ are
then restricted to the discrete set
\begin{equation}
	k=\frac{2\pi n}{L}
\end{equation}
with a positive integer $n$. The Hamiltonian for discrete modes can be derived
by following the previous steps but replacing $\int {\rm d}k$ with
$(2\pi/L)\sum_n$, $\delta(k-l)$ with $(L/2\pi)\delta_{kl}$,
$\tilde{\psi}(k(n))$ with $\sqrt{L/(2\pi)}\psi_n$.

We should also adjust the $\phi$-Hamiltonian to its minisuperspace
form. Starting with the original action (\ref{S}) and introducing homogeneity
such that $\frac{1}{2}\int{\rm d}t {\rm d}x \dot{\phi}^2= \frac{1}{2}L\int{\rm
  d}t \dot{\phi}^2$, we see that the minisuperspace momentum is given by
$\Pi_{\phi}= L\dot{\phi}$ and depends on $L$. The $\phi$-Hamiltonian of the
minisuperspace contribution therefore differs from (\ref{Hdens}) in that
$\Pi_{\phi}^2$ is replaced by $\Pi_{\phi}^2/L^2$. The Hamiltonian operator
that combines a minisuperspace $\phi$-contribution with a discrete set of
$\psi$-oscillators is then
\begin{equation} \label{Hmini}
	\hat{H}=\frac{\Pi_{\phi}^2}{2L}+LV(\phi)+
	\frac{1}{2}\sum_{n= 1}^{\infty}\left(\hat{\Pi}_{A,n}^2+\hat{\Pi}_{B,n}^2
	+\omega_{\phi}(k(n))^2(\hat{A}_n^2+\hat{B}_n^2)\right)+
\frac{1}{2}(\hat{\Pi}_{A,0}^2+  \omega_{\phi}(0)^{2}\hat{A}_0^2)\,. 
\end{equation}
Compared with a continuum of modes, we have to be careful with $n=0$ because
the zero mode $\psi(0)$ is real and therefore implies only one oscillator,
$A_0$.

\subsection{Effective mode equations}

As before, the Hamilton operator $\hat{H}$ implies a quantum Hamiltonian
$H_{\rm Q}=\langle\hat{H}\rangle$ evaluated in a generic state. We evaluate
this Hamiltonian to second semiclassical order and, in a first step, ignore
all cross-correlations. While this assumption constitutes a restriction on the
class of states that can be studied with the model, we will show that it is
self-consistent.  The assumption relies on the condition that the $k$-modes of
$\psi$ are initially decoupled in terms of moments or cross-correlations, and
remain so throughout evolution.

The modes are decoupled in our classical Hamiltonian.  A sufficient condition
for self-consistency of our assumption at the semiclassical level is then that
all quadratic moments that involve different $k$s remain zero if they vanish
in an initial state. The relevant equations of motion are obtained from
Poisson brackets, derived from (\ref{Poisson}), of the form
$\{\Delta(s_{n_1}s_{n_2}),\Delta_H\}$ where $\Delta_H$ is a moments that
appears in the quantum Hamiltonian $\langle\hat{H}\rangle$. 

For a second-order
expansion of $\langle\hat{H}\rangle$ with an $H$ free of classical
interactions between the modes, any $\Delta_H$ is of the form of either
$\Delta(\Pi_N^2)$ or $\Delta(s_N^2)$ where each $\Pi_N$ or $s_N$ refers to a
single mode. In this case, we have
\begin{equation}\label{CrossPoisson}
 \{\Delta(s_{n_1}s_{n_2}),\Delta(\Pi_N^2)\}=
 2\Delta(s_{n_1}\Pi_N)\delta_{n_2N}+ 2\Delta(s_{n_2}\Pi_N) \delta_{n_1N}\,.
\end{equation}
For a cross-covariance $\Delta(s_{n_1}s_{n_2})$, we have
$n_1\not=n_2$. Therefore, any moment that may appear on the right-hand side of
(\ref{CrossPoisson}) with a non-zero coefficient is a
cross-covariance. Analogous arguments hold for cross-covariances
$\Delta(s_{n_1}\Pi_{n_2})$ and $\Delta(\Pi_{n_1}\Pi_{n_2})$ of different
modes.  In general, therefore, calling the set of these mixed moments of
$k$-modes ${\cal M}$, Hamilton's equations generated by
$\langle\hat{H}\rangle$ using the Poisson bracket for moments are necessarily
of the form $\{m,\langle\hat{H}\rangle\}\propto \sum_{m'\in{\cal M}} a_{m'}m'$
for any $m\in{\cal M}$, with moment-independent coefficients
$a_{m'}$. Therefore, if all $m\in {\cal M}$ vanish initially, they remain zero
at all times in this model.

Moments of the state, on which $H_{\rm Q}$ depends, can therefore
self-consistently be expressed in canonical Darboux variables by using the
same mapping (\ref{sps}) known for a single degree of freedom, but applied
independently to each mode. This procedure leads to
\begin{eqnarray}
	H_{\rm Q}&= & \frac{\Pi_{\phi}^2}{2L}+LV(\phi)
	+\frac{1}{2}\sum_{n= 1}^{\infty}\left(\Pi_{A,n}^2+\Pi_{B,n}^2
	+\omega_{\phi}(k(n))^2(A_n^2+B_n^2)\right)+
\frac{1}{2}(\Pi_{A,0}^2+\omega_{\phi}(0)^{2}A_0^2)\nonumber\\ 
	&&
        +\frac{1}{2}\sum_{n=1}^{\infty}\left(p_{A,n}^2+p_{B,n}^2+
 \frac{U_{A,n}}{s_{A,n}^2} 
	+\frac{U_{B,n}}{s_{B,n}^2}+\omega_{\phi}(k(n))^2(s_{A,n}^2+s_{B,n}^2)\right)
      \label{HQfield}\\
	&& +\frac{1}{2}\left(p_{A,0}^2+\frac{U_{A,0}}{s_{A,0}^2}
 +\omega_{\phi}(0)^2s_{A,0}^2\right)+O(\hbar^{3/2}) \nonumber
\end{eqnarray}
with canonical quantum degrees of freedom $(s_{A,n},p_{A,n})$,
$(s_{B,n},p_{B,n})$, $U_{A,n}$ and $U_{B,n}$ such that
\begin{eqnarray}
	\Delta(A_n^2)&= & s_{A,n}^2 \quad,\quad 
	\Delta(\Pi_{A,n}^2)= p_{A,n}^2+\frac{U_{A,n}}{s_{A,n}^2}\\
	\Delta(B_n^2)&= & s_{B,n}^2 \quad,\quad 
	\Delta(\Pi_{B,n}^2)= p_{B,n}^2+\frac{U_{B,n}}{s_{B,n}^2}\,.
\end{eqnarray}
All other variables in (\ref{HQfield}) are understood
as expectation values of the basic mode, taken in the same state in which
moments are computed.

Using canonical Poisson brackets for all variables in (\ref{HQfield}) except
for the constant $U_{A,n}$ and $U_{B,n}$, we derive second-order equations of
motion 
\begin{eqnarray}
	\ddot{\phi}+V'(\phi)+\frac{\lambda\phi}{L}
 \left(\sum_{n=1}^{\infty}(A_{k}^2+B_{k}^2+s^{2}_{A,k}+s^{2}_{B,k})
+A_0^2+s_{A,0}^2\right)&=& 0 \label{phi-eq}\\
	\ddot{A}_{n}+\omega^2_{\phi}(k(n))A_n= 0 \quad\mbox{and}\quad
	\ddot{B}_{n}+\omega^2_{\phi}(k(n))B_n&=& 0 \mbox{ ($n>0$)} \\
	\ddot{s}_{A,n}-\frac{U_{A,n}}{s^3_{A,n}}+\omega^2_{\phi}(k(n))s_{A,n}= 0
        \quad\mbox{and}\quad
	\ddot{s}_{B,n}-\frac{U_{B,n}}{s^3_{B,n}}+\omega^2_{\phi}(k(n))s_{B,n}&=&
        0   \mbox{ ($n>0$)}\\
	\ddot{A}_0+\omega_{\phi}^2(k(0))A_0&=& 0\,. \label{A0-eq}
\end{eqnarray}
In the first line, we have used the specific frequency (\ref{Omega}).

\begin{figure}
	\centering
	\includegraphics[width=\textwidth]{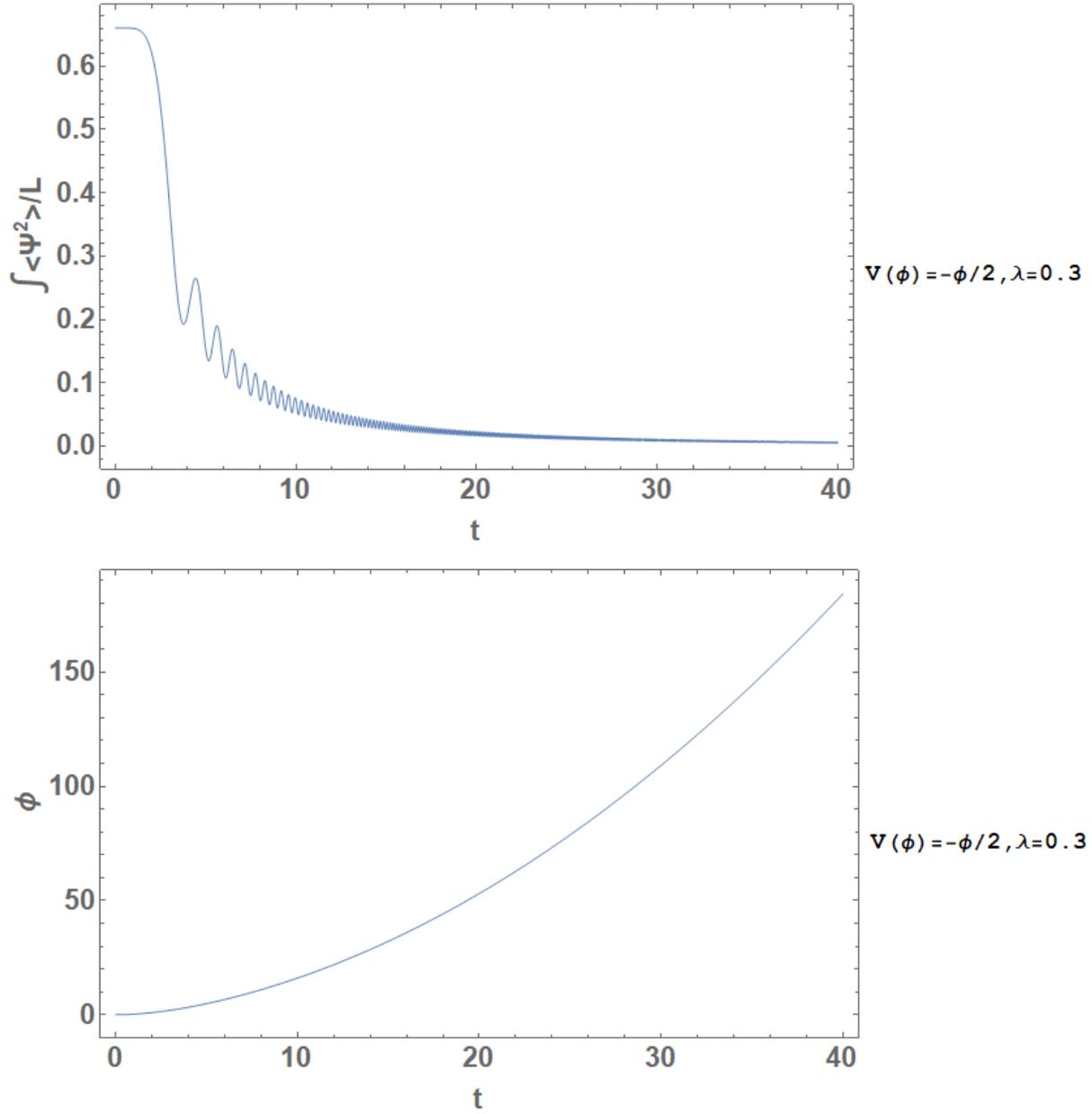}
	\caption{The backreaction term $\langle\hat{\psi}^2\rangle$ and
          background evolution $\phi(t)$ as functions of $t$, using
          $\lambda=0.3$. }
	\label{lambda-03}
\end{figure}

\begin{figure}
	\centering
	\includegraphics[width=\textwidth]{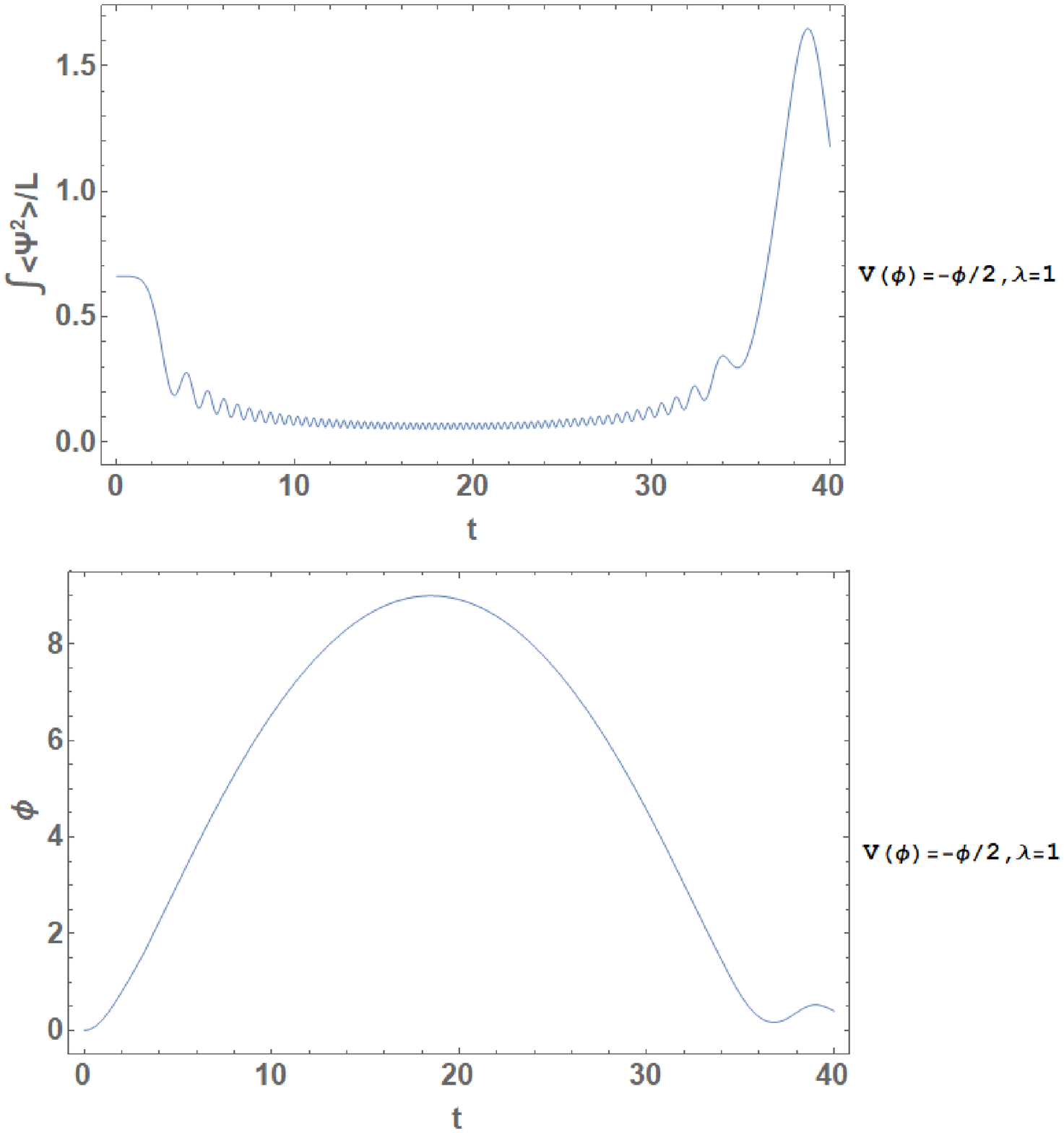}
	\caption{The backreaction term $\langle\hat{\psi}^2\rangle$ and
          background evolution $\phi(t)$ as functions of $t$, using
          $\lambda=1.0$. With this value, compared with Fig.~\ref{lambda-03},
          back-reaction is string enough to turn around $\phi$ before it grows
          large.}
	\label{lambda-10}
\end{figure}

Equations~(\ref{phi-eq})--(\ref{A0-eq}) are coupled and hard to solve
analytically, but numerical solutions can be obtained for specific initial
values. We assume a background potential $V(\phi)=-\frac{1}{2}\phi$ in what
follows and choose the moments of each $k$-mode to correspond to the Gaussian
ground state inititally, with frequency
$\omega_{\phi}(k)|_{t=0}=\sqrt{k^2+m^2+\lambda\phi(0)^2}$. In particular,
$U_{A,n}=U_{B,n}=\hbar^2/4$ for all $n$. The initial value for $\phi$ that
appears in the frequencies is assumed to vanish, as are all other dynamical
variables.

\begin{figure}
	\centering
	\includegraphics[width=0.9\textwidth]{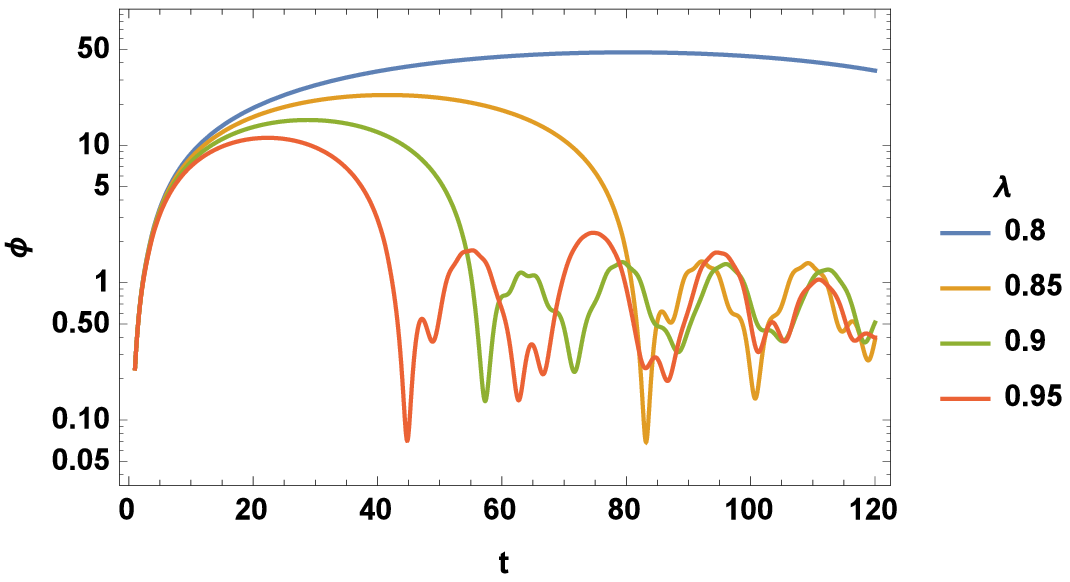}
	\caption{Background evolution $\phi(t)$ as functions of $t$ for
          various values of $\lambda$. The turn-around of $\phi$ is delayed
          for smaller $\lambda$, implying weaker back-reaction. }
	\label{log-plot}
\end{figure}

Figures~\ref{lambda-03} and \ref{lambda-10} show the background evolution
$\phi(t)$ and the magnitude $\int {\rm d}x\langle\hat{\psi}^2\rangle\approx
\sum_{k> 0}(A_{k}^2+B_{k}^2+s_{A,k}^2+s_{B,k}) +A_0^2+s_{A,0}^2$ of
back-reaction, using a momentum cutoff of $k_{\Lambda}=50\times 2\pi/L$ and
the parameters $m=0.1$, $L=100$, $\lambda=0.3$. Except for the momentum
cutoff, these parameters match the ones used in Figures 3--5 of
\cite{CQCFieldsHom}. Their momentum cutoff is $k_{\Lambda}'=4$ while for us it
is $k_{\Lambda}=3.96$. There are also differences in the treatment of quantum
fields, which explains why numerical evolutions in these two approaches do not
align precisely in quantitative terms. In addition, \cite{CQCFieldsHom} also
considers non-linear potentials $V(\phi)$ in detail, from which we refrain
here in a first analysis. For smaller $\lambda$, it takes longer and longer for $\phi$ to turn around as
a consequence of back-reaction. The longer phase of increasing $\phi$ makes it
difficult to resolve the turn-around numerically for very small $\lambda$, but
the general trend of a delayed turn-around is illustrated in
Figure~\ref{log-plot}.

Nevertheless, our numerical results are qualitatively comparable with those of
\cite{CQCFieldsHom}. The models are not identical because we evolve mode
equations on a compact space, while \cite{CQCFieldsHom} considers a lattice
approximation to evolve spatial fields. The homogeneity assumption on the
background is shared by both approaches, except for a small excursion into
inhomogeneous backgrounds at the end of \cite{CQCFieldsHom}. The underlying
equations are also identical because equation (2) of \cite{CQCFieldsHom},
given by
\begin{equation}
	\Box\phi+V'(\phi)+\lambda\langle\hat{\psi}^2\rangle\phi=0 \,,
\end{equation}
is equivalent to our mode equations for a homogeneous background. In
particular, our mode equations correspond to the Klein--Gordon equation
\begin{equation}
  \Box\phi+V'(\phi)+\frac{\lambda}{L}\int{\rm
    d}x\langle\hat{\psi}^2\rangle\phi=0
\end{equation}
with a source term for back-reaction. This equation is the same as equation
(9) of \cite{CQCFieldsHom} for a homogeneous $\phi$.

\subsection{Interactions and correlations with the background}

The homogeneity problem in inflationary cosmology is concerned with the
question of how it can be possible for inhomogeneity to develop out of an
initial homogeneous vacuum state, subject to translation invariant dynamics
that should not break this symmetry. (See for instance
\cite{Halliwell2,Halliwell3,InflStruc,CollapseStruc1,CollapseStruc2,CollapseStruc3}.)
Effective equations for moments of a quantum field or their canonical
variables can be a powerful tool to determine conditions under which
inhomogeneity may build up. Similar considerations show that quantum
correlations with the background may be relevant in certain
situations. Specific details of such back-reaction effects may also be
relevant for observational questions in cosmological eras at later times
\cite{BackreactionCosConst,BackreactionSuperHubble,BackreactionStochastic,BackreactionPreheating}.

\subsubsection{Inhomogeneity from a back-reacting vacuum}

We assume that our initial state for quantum field and classical background is
spatially homogeneous.  Suitable conditions on initial moments are expressed
in our notation as
\begin{eqnarray} \label{Symm}
	&&\langle\psi_k\rangle= 0 \quad\mbox{and
          therefore}\quad\langle\psi\rangle=0\quad\mbox{when}\quad t=0\\ 
	&&\Delta(\psi_k\psi_{l\neq k})=
        \Delta(\pi_i\psi_j)=0\quad\mbox{when}\quad t=0\\ 
 	&&\phi= \phi_0(t)+\sum_n \phi_n(t)e^{ik_nx}\quad\mbox{with}\quad
        \phi_n(0)=0\,, 
\end{eqnarray}
where the sum excludes $n=0$, or alternatively $\phi_0(t)=0$ for all
$t$. 

At this point it is important to note that choosing a specific initial state
poses a condition on initial values for evolution, but it should not restrict
the general dynamics of a background interacting with a quantum
field. Therefore, in order to consider the possibility of background
inhomogeneity being generated, the Hamiltonian should contain kinetic,
potential, and interaction terms for a generic inhomogeneous background field
$\phi(x,t)$, including its momentum $\Pi(x,t)$ in a canonical formulation. If
we were to represent the background as a homogeneous minisuperspace model as
in (\ref{Hmini}), background inhomogeneity would be excluded by fiat. In order
to see how inhomogeneity may be generated, we should start with the more
general Hamiltonian (\ref{Hphi}) with a field Hamiltonian density ${\cal
  H}_{\phi}$ and study the evolution it generates starting with symmetric
initial conditions (\ref{Symm}) for a homogeneous initial state. The
expectation value $H_{\rm eff}=\langle\hat{H}\rangle$ in a generic field state
is a suitable effective Hamiltonian for this analysis. Here we encounter the
important distinction between quantum field theory on a homogeneous
background, which would be described by our minisuperspace background, and a
more complete treatment of an interacting field and background.

A key contribution in this Hamiltonian that does not appear in the
minisuperspace version is the interaction term
\begin{equation}
\int {\rm d}x \phi^2\langle\hat{\psi}^2\rangle=\sum_{N}\left(
  \phi_0^2\Delta(\psi_{k_N}\psi_{-k_N}) 
	+2\phi_0\sum_n\phi_n\Delta(\psi_{k_N}\psi_{-k_N-k_n})+\sum_{n,l}
	\phi_n\phi_l
	\Delta(\psi_{k_N}\psi_{-k_N-k_n-k_l})\right)\,.
\end{equation}
Using symmetric initial conditions, the momentum $\Pi_m$ conjugate to
$\phi_m$ evolves according to
\begin{eqnarray}
	\dot{\Pi}_m&=& \{\Pi_m,H_{\rm eff}\}
	\sim \{\Pi_m, \int{\rm d}x\phi^2\langle \hat{\psi}^2\rangle\}\nonumber\\
	&=& \sum_N \{\Pi_m,\phi_0^2\Delta(\psi_{k_N}\psi_{-k_N})
	+2\phi_0\sum_n\phi_n\Delta(\psi_{k_N}\psi_{-k_N-k_n})+\sum_{n,l}
	\phi_n\phi_l
	\Delta(\psi_{k_N}\psi_{-k_N-k_n-k_l})\}\nonumber\\
	&=& 2\sum_N \phi_0\Delta(\psi_{k_N}\psi_{-k_N-k_m})\,,
\end{eqnarray}
keeping only terms that have a chance of being non-zero.  Given our initial
conditions, the last integrand is initially non-zero if and only if
$k_N=-k_m/2$, and therefore $\dot{\Pi}_m\sim
\phi_0\Delta(\psi_{-k_{m/2}}\psi_{-k_{m/2}})$ does not vanish.  (Physically,
$\langle \psi^2\rangle$ cannot be perfectly homogeneous due to the uncertainty
relation, such that $\Delta(\psi_{k_N}\psi_{k_N})\neq 0$.) 

Even for an initially homogeneous $\phi$, the coupling $\phi^2\langle
\psi^2\rangle$ implies that the background field $\phi$, seen as a harmonic
oscillator, may have a position-dependent frequency $\langle
\psi^2\rangle$. If the initial state is completely homogeneous, the frequency
$\langle \psi^2\rangle$ is homogeneous and no background inhomogeneity is
generated, in accordance with the expectation based on translation-invariant
dynamics. However, this condition on the initial state is rather strong
because it imposes homogeneity in all of space at an initial time, potentially
far beyond the Hubble radius that would be accessible by causality. This
assumption is not consistent with the generic behavior of space-time expected
near a spacelike singularity such as the big bang according to the
Belinskii--Khalatnikov--Lifshitz (BKL) scenario \cite{BKL}. 

There should therefore be a certain mode number at which strict homogeneity
breaks down. On these scales, the background field oscillates differently at
different places and cannot stay homogeneous. One might expect that such
large-scale inhomogeneity far outside the Hubble radius may not be relevant,
but we will now show that moment dynamics of the field $\psi$ implies that
large-scale inhomogeneity eventually trickles down to smaller scales, at which
the background inhomogeneity it generates will be relevant.

It is perhaps surprising that moment terms from a quantum field allow
inhomogeneity to travel from large scales into a cosmological horizon. Our
results show that this is indeed possible even in a quantization of a
classically causal theory because we start with standard covariant action of a
scalar field. In fact, as our derivation shows, inhomogeneity builds up on
smaller scales through a gradual process of growing cross-correlations, rather
than direct propagation that would be impossible in a causal theory. 

A useful interpretation of the gradual process we derived is as an effect of
quantum non-locality. Moment methods make it possible to formulate effective
field theories without performing a derivative expansion. This feature has
been know for some time in an application to quantum mechanics, such as in
quantum chemistry \cite{QHDTunneling} where they give access to non-adiabatic
reaction dynamics. In \cite{EffAc,Karpacz,HigherTime}, it was shown explicitly
that a combination of moment methods with an adiabatic approximation yields
results equivalent to the low-energy effective potential \cite{EffAcQM} or
higher-derivative corrections. Similarly, in an application to scalar field
theories, a combination of moment methods and an adiabatic approximation
\cite{CW} allows one to rederive the Coleman--Weinberg potential
\cite{ColemanWeinberg}. In the present paper, however, we have applied moment
methods to quantum field theories without using an adiabatic
approximation. 

Instead of higher-derivative corrections, our effective Hamiltonians implement
quantum corrections through coupling terms to non-classical fields as new
independent degrees of freedom, such as the modes $s_{A,k}$ and $s_{B,k}$ in
canonical terms or $\Delta(\psi_k\psi_l)$ directly in terms of moments. It is
possible to interpret such an extended field theory as a local formulation of
a non-local theory without extra degrees of freedom: If one were able to solve
equations of motion for the new fields as functions of the classical fields
and to insert them in the extended Hamiltonian, local terms would be replaced
by integrals representing solutions of partial differential equations. It
would be cumbersome to do so in practice, and it is in fact easier to analyze
the local extended system. Nevertheless, the argument demonstrates that the
build-up of inhomogeneity derived here is a consequence of quantum
non-locality, a crucial new ingredient in early-universe cosmological models.

\subsubsection{Field correlations}

If inhomogeneity is included in the background Hamiltonian and its coupling
terms, new interaction channels open up for oscillator moments that are not
contained in models with a strictly homogeneous background. To simplify the
discussion, we assume now that the background is homogeneous up to adding a
single mode, $\phi(t,x)=\phi_0(t)+ \phi_K(t) e^{iKx}$. The interaction
term then depends on the $\phi$-modes through the terms $\phi_0
\phi_Ke^{iKx}$ and $\phi_K^2e^{2iKx}$ obtained directly from $\phi^2$. The
non-local expression (\ref{N}) is then again local, but with additional
interaction terms:
\begin{eqnarray}
&& \tilde{\psi}(k) \int_{-\infty}^{\infty}{\rm d}l  \int{\rm d}x
 \phi(t,x)^2 e^{i(l+k)x}\tilde{\psi}(l) \nonumber\\
&=& \tilde{\psi}(k)
 \left(\phi_0^2\int_{-\infty}^{\infty}{\rm d}l 
   \int{\rm d}x 
  e^{i(l+k)x}\tilde{\psi}(l)+ 2\phi_0\phi_K \int_{-\infty}^{\infty}{\rm d}l 
   \int{\rm d}x 
  e^{i(K+l+k)x}\tilde{\psi}(l)\right.\nonumber\\
&&\qquad\left.+ \phi_K^2 \int_{-\infty}^{\infty}{\rm d}l 
   \int{\rm d}x 
  e^{i(2K+l+k)x}\tilde{\psi}(l)\right)\nonumber\\
&=& \phi_0^2 \tilde{\psi}(k)\tilde{\psi}(-k) + 2\phi_0\phi_K \tilde{\psi}(k)
\tilde{\psi}(-k-K)+ 
  \phi_K^2 \tilde{\psi}(k)\tilde{\psi}(-k-2K) \label{Ninhom}
\,. 
\end{eqnarray} 

There are therefore interactions between neighboring modes over distances $K$
and $2K$. Any interaction potential of the form
$V(\tilde{\psi}(k),\tilde{\psi}(k+k_0))$ with some constant $k_0$ implies
evolving cross-correlations, which we express here directly in terms of
moments without a canonical parameterization. A second-order effective
Hamiltonian contains the moments
$\Delta(\tilde{\psi}(k)\tilde{\psi}(k+k_0))$. Since the variances
$\Delta(\tilde{\psi}(k)^2)$ cannot be zero as per uncertainty relations, the
effective Hamiltonian implies non-zero time derivatives of moments of the form
$\Delta(\tilde{\psi}(k+k_0)\tilde{\Pi}_{\psi}(k))$, using the Poisson bracket
\begin{equation}
 \{\Delta(\tilde{\psi}(k+k_0)\tilde{\Pi}_{\psi}(k)),
\Delta(\tilde{\psi}(k)\tilde{\psi}(k+k_0))\}=- 
\Delta(\tilde{\psi}(k+k_0)^2)\,.
\end{equation}
Similarly, $\Delta(\tilde{\psi}(k-k_0)\tilde{\Pi}_{\psi}(k))$ has a non-zero time
derivative because
\begin{equation}
 \{\Delta(\tilde{\psi}(k-k_0)\tilde{\Pi}_{\psi}(k)),
\Delta(\tilde{\psi}(k)\tilde{\psi}(k-k_0))\}=- 
\Delta(\tilde{\psi}(k-k_0)^2)\not=0\,.
\end{equation}
Once $\Delta(\tilde{\psi}(k\pm k_0)\tilde{\Pi}_{\psi}(k))$ have evolved to
non-zero values (assuming they start at zero for an uncorrelated initial
state), the time derivatives of $\Delta(\tilde{\psi}(k)\tilde{\psi}(k\pm
k_0))$ will be non-zero, using
\begin{equation}
  \{\Delta(\tilde{\psi}(k)\tilde{\psi}(k\pm k_0),
(\Delta(\tilde{\Pi}_{\psi}(k)^2\}=2
  \Delta(\tilde{\psi}(k\pm k_0)\tilde{\Pi}_{\psi}(k))
\end{equation}
and a contribution from the kinetic energy to the effective Hamiltonian.

The state therefore evolves to non-zero correlations at a $k$-space distance
$k_0$. Then, we have non-zero time derivatives of
$\Delta(\tilde{\psi}(k+2k_0)\tilde{\Pi}_{\psi}(k))$ because of a non-zero
\begin{equation}
  \{\Delta(\tilde{\psi}(k+2k_0)\tilde{\Pi}_{\psi}(k)),
\Delta(\tilde{\psi}(k)\tilde{\psi}(k+k_0))\}=- 
  \Delta(\tilde{\psi}(k+2k_0)\tilde{\psi}(k+k_0))\not=0
\end{equation}
where $\Delta(\tilde{\psi}(k)\tilde{\psi}(k+k_0))$ is again a term in the
effective potential. Non-zero
$\Delta(\tilde{\psi}(k+2k_0)\tilde{\Pi}_{\psi}(k))$ then imply non-zero time
derivatives of $\Delta(\tilde{\psi}(k)\tilde{\psi}(k+2k_0))$ through the
kinetic terms of the effective Hamiltonian. Iteration shows that shorter-range
correlations of increasing $k$ in $\psi$ develop over time, provided the
background is not exactly homogeneous.

Turning now to evolution equations for moments of background modes of
inhomogeneity, the middle term in (\ref{Ninhom}), inserted in an effective
Hamiltonian, shows that any momentum $\Pi_{\phi}(K)$ of a $\phi$-mode, such
that $\Delta(\tilde{\psi}(k)\tilde{\psi}(-k-K))$ has developed a non-zero
value, has a non-zero time derivative. Even if the initial $\psi_K$ was zero,
therefore, background inhomogeneity develops in a large number of modes (all
integer multiples of $K$) provided there was some initial inhomogeneity in
just one mode, $K$. If the wave length corresponding to the initial $K$ is
super-Hubble, as suggested by the BKL scenario, correlations in $\psi$ and
then inhomogeneity in $\phi$ trickles down to smaller wave lengths (larger
$k$) that eventually enter the Hubble radius and become relevant for the
history of our visible universe. These results provide a consistent picture of
the generation of background inhomogeneity in cosmology. (We note that the
situation in cosmology is different from quantum phase transitions in which
structure may also form, as studied for instance in \cite{Kinks}. In the
latter case, an infinite-volume limit is required for a mathematical
description of the phase transition, and homogeneity of the pre-transition
state may be assumed even in the limit. In cosmology, the BKL scenario
prevents one from making the same assumption.)

\subsubsection{Background correlations in quantum field theories}

In addition to indirect effects of field correlations on the background
dynamics as just discussed, a general quantum description of background and
field modes may contain direct quantum correlations between the background and
the field. Such correlations are also excluded by fiat in a strict background
treatment, but they can be included in an extended model using our new
methods. We briefly present here such models, but for now refrain from
analyzing them.

The restricted second-order Hamiltonian (\ref{HCorr}) suggests an extension of
the background-field model to multiple canonical fields for background and
oscillators. Some of the new fields represent quantum degrees of freedom for
fluctuations, as before, and others cross-correlations between background and
oscillator fields. Promoting all variables in (\ref{HCorr}) to fields, we
obtain the Hamiltonian density
\begin{eqnarray} \label{HCorr2}
 	\mathcal{H}&=&\frac{1}{2}\left(\Pi_{\phi}^2+\Pi_{\phi_1}^2+\Pi_{\phi_2}^2
+(\partial_x\phi)^2 +(\partial_x\phi_1)^2
+(\partial_x\phi_2)^2\right)  
+V(\phi) \nonumber\\
&&+\frac{1}{2}\left(\Pi_{\psi}^2+\Pi_{\psi_1}^2+\Pi_{\psi_2}^2+\Pi_{\psi_3}^2+
(\partial_x\psi)^2+(\partial_x\psi_1)^2+(\partial_x\psi_2)^2+(\partial_x\psi_3)^2
\right)\\
&&+\Omega_{\phi}(t,x)^2\left(\psi^2+\psi_1^2+\psi_2^2+\psi_3^2\right) +
\frac{1}{2}\lambda (\phi_1^2+\phi_2^2) \psi^2+ 2\lambda
\phi\sqrt{\phi_1^2+\phi_2^2} \psi \psi_3 \nonumber
\end{eqnarray}
with five new fields, $\phi_1$, $\phi_2$, $\psi_1$, $\psi_2$ and $\psi_3$, and
the same $\Omega_{\phi}(t,x)$ as before.  The field $\psi_3$ describes
background correlations.

For a homogeneous background
$(\phi,\phi_1,\phi_2)$, a mode expansion is still possible, resulting in 
local, decoupled Hamiltonians for the modes. The potential
\begin{eqnarray}
&& \Omega_{\phi}(t)^2\left(\psi^2+\psi_1^2+\psi_2^2+\psi_3^2\right) +
\frac{1}{2}\lambda (\phi_1^2+\phi_2^2) \psi^2+ 2\lambda
\phi\sqrt{\phi_1^2+\phi_2^2} \psi \psi_3\nonumber\\
&=& \Omega_1(t)^2 \psi^2+ \Omega_2(t)^2(\psi_1^2+\psi_2^2)+
\Omega_2(t)^2\psi_3^2+ \omega(t) \psi\psi_3
\end{eqnarray}
then implies three different frequencies,
\begin{equation}
 \Omega_1(t)^2 =
 \Omega_{\phi}(t)^2+\frac{1}{2}\lambda(\phi_1(t)^2+\phi_2(t)^2)
\end{equation}
for $\psi$ and
\begin{equation}
 \Omega_2(t)^2= \Omega_{\phi}(t)^2
\end{equation}
for $\sqrt{\phi_1^2+\phi_2^2}$ as well as $\psi_3$, and a rotation coefficient
\begin{equation}
 \omega(t)= 2\lambda \phi(t)\sqrt{\phi_1(t)^2+\phi_2(t)^2}
\end{equation}
between $\phi$ and $\phi_3$. As before, beat-like effects are expected by mode
mixing. 

If the background is homogeneous, a single correlation field $\psi_3$ is
sufficient to extend the model to background correlations. In addition, there
may be correlations between different modes of the field $\psi$, which at
present is hard to describe in canonical quasiclassical form because a
complete set of Casimir--Darboux for a quantum field remains unknown. Instead,
one may construct a completely correlated background-field model by keeping
the background correlation field $\psi_3$ in canonical quasiclassical form,
while fully quantizing the field $\psi$ or its modes as before. In
(\ref{HCorr2}), $\psi$, $\psi_1$ and $\psi_2$ as well as their momenta would
then be replaced by a single field operator $\psi$, while $\psi_3$ remains a
single field of classical type. In this formulation, hybrid methods of
classical-quantum dynamics, such as
\cite{ClassQuantMeas,HybridChem,HybridQuant,HybridMadelung}, would find a
useful place in cosmology.

\section{Conclusions}

We have significantly extended a back-reaction model considered in \cite{CQC}
in order to study implications of backreaction and the build-up of
inhomogeneity in cosmological evolution. To this end, we first showed that the
formalism of \cite{CQC} is equivalent to a special case of non-adiabatic
semiclassical approximations obtained from the dynamics of moments
parameterizing an evolving state. The restricted nature of the formalism
developed in \cite{CQC} implies that there are key distinctions between the
two approaches that provide different advantages, depending on what physical
system is being studied.

The formalism of \cite{CQC} is efficient in tackling quadratic Hamiltonians
and Gaussian approximations. For anharmonic systems, its approximate equations
of motion remain quite simple, but there is no self-consistent way to
determine whether the approximation is reliable. This shortcoming is related
to a lack of physical interpretation of the degrees of freedom, $\xi$ and
$\chi$, used in \cite{CQC} to describe quantum variables. The derivation of
how these variables appear in a Hamiltonian or in equations of motion requires
a wave function $\psi$, usually assumed Gaussian, but the reduction of
infinitely many quantum degrees of freedom implicitly described by the
functional dependence of $\psi$ to just two relevant ones is not systematic.
In cases in which no suitable wave function is known, for instance in
situations of particle creation that lead one away from a vacuum state, the
predictability of the formalism remains unclear. No recipe for going beyond
leading order (in some kind of loop expansion) has been developed.

Our embedding of the formalism of \cite{CQC} in the systematic framework of
canonical effective theory helps to make the approximations much more
systematic. The quantum degrees of freedom are now physically interpreted as
moments of a state, not just in Gaussian or near-Gaussian situations. There is
a clear extension to higher orders of moments, mimicking the loop expansion of
interacting theories. To low orders, as we have shown, explicit canonical
realizations are available and can be used to extend the model of \cite{CQC}
to correlation degrees of freedom. As we have argued, these degrees of freedom
are crucial for cosmological back-reaction in which no sharp physical
separation between background and perturbations exists, owing to general
covariance. Correlation degrees of freedom are then seen to give rise to new
beat-like effects that may be relevant in particle production.

Extended models also show how background inhomogeneity may be generated out of
a symmetric sub-horizon initial state, and what role may be played by various
forms of quantum correlations. We recognized that the extension to moments as
independent degrees of freedom can be interpreted as a local formulation of
quantum non-locality, which in this form has not been considered before in
early-universe models. By an explicit calculation of equations of motion for
moments we showed that quantum non-locality implies a gradual build-up of
inhomogeneity within a cosmological horizon, even if the initial sub-horizon
state is completely homogeneous but inhomogeneity exists on much larger
scales. This build-up of inhomogeneity is a consequence of quantum
non-locality and does not violate causal propagation. The new effect is
therefore able to solve the homogeneity problem of inflationary structure
formation.

\section*{Acknowledgements}

This work was supported in part by NSF grant PHY-1912168. We thank Mainak
Mukhopadhyay and Tanmay Vachaspati for discussions and useful suggestions.

%\bibliographystyle{../preprint}
%\bibliography{../Bib/QuantGra,../Bib/Tunneling}

\end{document}